\documentclass[11pt,a4paper]{article}
\usepackage{jheppub}
\usepackage{mathtools}
\usepackage{mathrsfs}
\usepackage{amsmath}
\usepackage{tikz-feynman}
\usepackage{amssymb}
\usepackage{slashed}
\usepackage{bbm}
\usepackage{bm}
\usepackage{tikz}
\usepackage{natbib}
\usepackage{float}
\usepackage[]{todonotes}
\usepackage{dsfont}
\usepackage{hyperref}
\usepackage{braket}
\usepackage{amsmath,epsf,amssymb,latexsym,amsthm,setspace,array,pifont,color,hyperref,amsfonts,dsfont,cancel,braket,parskip}
\usepackage[none]{hyphenat} % No hyphenated line endings
\usepackage{graphicx} % Required for including pictures, diagrams
\def\pd{\partial}
\def\D{\Delta}

\def\C2{{C_{ES^2}}}
\def\CB3{{C_{BS^3}}}

\def\vS1{\bm S_1}

\def\cE{\mathcal{E}}

\newcommand{\sd}{\mathrm{d}}

\newcommand{\dd}{\hat{\mathrm{d}}}

\newcommand{\bb}[1]{\mathbb{#1}}
\newcommand{\cl}[1]{\mathcal{#1}}

\def\spvecA#1;{\if;#1;\else #1\cr \expandafter \spvecA \fi}
\usepackage{tcolorbox}
\definecolor{airforceblue}{rgb}{0.36, 0.54, 0.66}
\definecolor{azure}{rgb}{0.0, 0.5, 1.0}
\newtcolorbox{tdbox}{colback=airforceblue!40!white,colframe=azure!90!black} 
\newcommand{\td}[1]{
	\if\notesOn1
	\begin{tdbox}
		#1
	\end{tdbox}
	\fi
}

\def\notesOn{1}
\tikzset{
	graviton/.style={
		double,
		decoration={snake, aspect=0.75, mirror, segment length=1.5mm},
		decorate
	}
}

\tikzfeynmanset{
	bigblob/.style={
		shape=circle,
		draw=blue,
		fill=red}
}
\renewcommand{\[}{\begin{equation}\begin{aligned}}
		\renewcommand{\]}{\end{aligned}\end{equation}}
\renewcommand{\[}{\begin{equation}\begin{aligned}}
		\renewcommand{\]}{\end{aligned}\end{equation}}

\renewcommand{\d}{\mathrm{d}}
\newcommand{\del}{\hat{\delta}}

\newcommand{\amp}{\mathcal{A}}

\def\Lexp{\biggl\langle\!\!\!\biggl\langle}
\def\Rexp{\biggr\rangle\!\!\!\biggr\rangle}

\def\Lexp{\biggl\langle\!\!\!\biggl\langle}
\def\Rexp{\biggr\rangle\!\!\!\biggr\rangle}

\definecolor{emerald}{rgb}{0.31, 0.78, 0.47}

\definecolor{allOrderBlue}{rgb}{0.4,0.5,1}
\title{Observables from the Spinning Eikonal}
\author[1]{Andres Luna,}
\author[2,3,4]{Nathan Moynihan,}
\author[3]{Donal O'Connell,}
\author[3]{Alasdair Ross,}
\affiliation[1]{Niels Bohr International Academy,
	Niels Bohr Institute, University of Copenhagen,
	Blegdamsvej 17, DK-2100, Copenhagen \O , Denmark}
\affiliation[2]{School of Mathematics \& Hamilton Mathematics Institute, Trinity College Dublin, College Green, Dublin 2, Ireland}
\affiliation[3]{Higgs Centre for Theoretical Physics, School of Physics and Astronomy, The University of Edinburgh, EH9 3FD, Scotland}
\affiliation[4]{Centre for Theoretical Physics, School of Physical and Chemical Sciences, Queen Mary University of London, 327 Mile End Road, London E1 4NS, United Kingdom}
\abstract{
We study the classical dynamics of spinning particles using scattering amplitudes and eikonal exponentiation.
We show that observables are determined by a simple algorithm.
A wealth of complexity arises in perturbation theory as positions, momenta and spins must be iteratively corrected at each order.
Even though we restrict ourselves to one-loop computations at quadratic order in spin, nevertheless we encounter and resolve a number of subtle effects.
Finally, we clarify the links between our work and various other eikonal approaches to spinning observables.
}

\begin{document}
	
	\maketitle
	
	\section{Introduction}

The direct detection of gravitational waves by the LIGO/Virgo collaboration~\cite{LIGOScientific:2016aoc,  LIGOScientific:2017vwq} launched a new era in multi-messenger astronomy, driving a demand for high-precision predictions for binary dynamics in general relativity (GR). 
In this context, the connection between scattering amplitudes in quantum field theory and the two-body problem in GR~\cite{ Damour:2016gwp,   Damour:2017zjx,   Bjerrum-Bohr:2018xdl} has seen increased interest. Indeed, in recent years, new systematic methods have been developed for extracting potentials and physical observables from
scattering amplitudes~\cite{Cheung:2018wkq, Kosower:2018adc,
Cristofoli:2019neg,   Bjerrum-Bohr:2019kec,   Brandhuber:2021eyq}. These aim to leverage modern methods for calculating amplitudes, including the double copy~\cite{  Bern:2019prr} and advanced integration techniques~\cite{Parra-Martinez:2020dzs}.   
By manifestly maintaining Lorentz invariance, these amplitudes-based approaches fit naturally in the post-Minkowskian (PM) framework, where observables are expanded  in Newton's constant $G$ while keeping their exact velocity dependence.
This extraction of classical physics from scattering amplitudes was used to produce the first conservative two-body Hamiltonian at $\mathcal O(G^3)$ and $\mathcal O(G^4)$~\cite{Bern:2019nnu,  Bern:2019crd,  Bern:2021dqo,  Bern:2021yeh} (see also Refs.~\cite{Kalin:2020fhe,  Dlapa:2021npj,  Dlapa:2021vgp, Dlapa:2022lmu,Bjerrum-Bohr:2022ows,Jakobsen:2023ndj,Jakobsen:2023hig} for results at these orders from alternative formulations). 

As gravitational-wave detectors become more
sensitive~\cite{Punturo:2010zz,  LISA:2017pwj,  Reitze:2019iox}, the
spin of the black holes or neutron stars in the binary will play a more important role in the interpretation of signals.  
Furthermore, adding spin  leads to more complex dynamics, since angular momentum can be exchanged between the bodies and the orbital motion, and the system is no longer confined to a plane.

There has also been swift development in the study of the dynamics of spinning objects interacting gravitationally within the PM approximation, as a result of the application of a variety of techniques including classical GR ~\cite{Bini:2017xzy, Vines:2017hyw,Bini:2018ywr} (see also Refs.~\cite{Damgaard:2022jem,Hoogeveen:2023bqa}). 
Later, a connection was established between Kerr black holes and a novel description of  scattering amplitudes for massive spinning fields in terms of the spinor helicity formalism ~\cite{Guevara:2017csg,Guevara:2018wpp,Chung:2018kqs,Moynihan:2019bor,Chung:2019yfs,Chung:2019duq,Guevara:2019fsj,Chung:2020rrz,Vines:2018gqi}. This fascinating relation has driven a quest to find scattering amplitudes which correctly describe the Kerr black hole to any order in spin, beyond linearised level in $G$. 
This question has been approached using massive higher-spin quantum field theories in Refs.~\cite{Chiodaroli:2021eug, Cangemi:2022abk, Cangemi:2022bew,Ochirov:2022nqz,Cangemi:2023ysz,Alessio:2023kgf}. Alternatively, Refs.~\cite{Bautista:2022wjf,Bautista:2023sdf} used solutions of the Teukolsky equation to derive the Compton amplitude (see Ref.~\cite{Bautista:2023szu} for a study of two-body dynamics derived from it).

More important for the purposes of this paper is the extraction of classical effects from scattering amplitudes. This has been achieved using diverse techniques, including spinning generalisations of effective field theories to produce Hamiltonians ~\cite{Bern:2020buy,Bern:2020uwk,Kosmopoulos:2021zoq,FebresCordero:2022jts,Bern:2022kto}, and of the observable-based method of Kosower, Maybee and O'Connell (KMOC) ~\cite{Maybee:2019jus,Menezes:2022tcs}. Another successful approach has been the use of heavy-particle effective theories~\cite{Damgaard:2019lfh,Aoude:2020onz,Aoude:2020ygw,Aoude:2021oqj,Chen:2021kxt,Chen:2022clh,Aoude:2022trd,Bjerrum-Bohr:2023jau, Haddad:2023ylx, Aoude:2023vdk,Bjerrum-Bohr:2023iey}, which directly target classical contributions.
Important effects like radiation and absorption have been studied in Refs. \cite{Brandhuber:2023hhl,DeAngelis:2023lvf} and ~\cite{Saketh:2022xjb, Aoude:2023fdm,Aoude:2023dui,Chen:2023qzo}, respectively.

The electromagnetic case, has been shown to be similar in structure~\cite{Saketh:2021sri, Bern:2021xze}, with the post-Lorentzian (PL) expansion being the  analog of the gravitational PM expansions, and so it has been used as a toy model to study spin effects~\cite{Arkani-Hamed:2019ymq, Kim:2023drc, Bern:2023ity} (see also Refs.~\cite{delaCruz:2020bbn, delaCruz:2021gjp} for non-abelian generalisations). 

In parallel to the program of Amplitudes, worldline-inspired effective field theories have been used with great success for high precision computations ~\cite{Liu:2021zxr,Jakobsen:2021lvp,Jakobsen:2021zvh,Jakobsen:2023ndj,Jakobsen:2023hig}  (see also Refs. ~\cite{Ben-Shahar:2023djm,Scheopner:2023rzp} for the derivation of classical amplitudes from the worldline).

In this paper, we focus on the \textit{eikonal approximation}, which has a long and successful history of obtaining classical gravitational physics from scattering amplitudes \cite{Ciafaloni:2018uwe,DiVecchia:2021bdo,Cristofoli:2021jas,Heissenberg:2021tzo,Emond:2021lfy,DiVecchia:2022nna,Haddad:2021znf,Adamo:2021rfq,DiVecchia:2022piu,Gatica:2023iws,Georgoudis:2023eke} --- see Ref. \cite{DiVecchia:2023frv} for a recent review. In particular, we follow-up on Ref.~\cite{Cristofoli:2021jas}, refining the techniques developed there to obtain spin effects in classical observables. In this approximation, the incoming energy is large compared to momentum transfers, and the amplitude can be separated into two factors. One of these factors is the exponent of an eikonal function which we denote as $\chi$, and we refer to it as the eikonal phase. This part of the amplitude encodes the classical physics. The remaining terms in the amplitude, which do not exponentiate, are quantum mechanical. The eikonal phase $\chi$ is analogous to a classical action, and so it can be used to derive classical observables at definite orders in Newton’s constant $G$, such as the scattering angle or the linear impulse.

We are interested in the eikonal phase related to four-point amplitudes in a quantum theory of massive, spinning fields. The exponentiation properties of the gravitational impact-parameter space amplitude in this setup have been investigated, for example in Refs. \cite{Adamo:2021rfq,Haddad:2021znf} (see Refs.~\cite{Heissenberg:2023uvo,Alessio:2022kwv,Bianchi:2023lrg} for related studies). Our main result is a simple procedure to derive classical observables, resulting in compact formulas for the impulse and the spin kick directly from the Eikonal phase.

We begin in section \ref{sect:Uncertainty} by introducing a translation operator in momentum space, and analysing its effect on the relation between the final state in a scattering process and the eikonal phase. We show that, in conjunction with the stationary phase conditions, it generates the relevant iterations to compute observables. In section \ref{sect:Observables}, we detail a prescription for obtaining observables from the eikonal phase in our setup, and in section \ref{sect:Comparisons} we compare to existing approaches in the literature.
Finally, in section \ref{section:Conclusions} we discuss our results and conclude.
We defer the presentation of results up to quadratic order in spin in the 2PM approximation order to appendix \ref{app:Observables}. We also show explicitly the Fourier transform and other useful formulas in appendix \ref{app:Useful}.

\textbf{Note added:} As this manuscript was in its final stages, Ref.~\cite{Gatica:2023iws} appeared on the arXiv, which contains some overlapping results.
 
	\section{Translation Operators and Eikonal exponentiation}
 \label{sect:Uncertainty}
	We begin with the assumption that eikonal exponentiation for spinning particles is of the form \cite{Cristofoli:2021jas}
	\[\label{eikonal_spin1}
		\exp\bigg(&{i\chi(\Pi(p_1,p_2)\cdot x,p_1,p_2)/\hbar}\bigg)^{a_1',a_2'}_{a_1,a_2} \\
  &\qquad= \delta^{a_1'}_{a_1}\delta^{a_2'}_{a_2}+ i \int \hat{\sd}^4q \, \del(2p_1\cdot q)\del(2p_2\cdot q)\,e^{-iq\cdot x/\hbar}\, \amp_4(p_1,p_2,q)^{a_1',a_2'}_{a_1,a_2} \,,
	\]
	where $\mathcal{A}_4(p_1,p_2,q)$ is the all-order $2\rightarrow 2$ amplitude with momentum transfer $q$, and we will ignore the quantum remainder.	
	The eikonal phase $\chi$ is a function of the momentum that determines the scattering plane, e.g. the plane orthogonal to $p_1,p_2$ above. We note especially that $\chi$ depends on $x$ via its projection onto this plane $\Pi(p_1,p_2)\cdot x$.
	
	Let's now consider a two-particle initial state with spin
	\[
	\ket{\Psi} & \equiv \int \mathrm{d} \Phi\left(p_1, p_2\right) \phi_b\left(p_1, p_2\right) \xi_1^{a_1} \xi_2^{a_2} a^{\dagger}\left(p_1\right)_{a_1} a^{\dagger}\left(p_2\right)_{a_2}|0\rangle \\
	& \equiv \int \mathrm{d} \Phi\left(p_1, p_2\right) \phi_b\left(p_1, p_2\right) \xi_1^{a_1} \xi_2^{a_2}\left|p_1, a_1 ; p_2, a_2\right\rangle \\
	& \equiv \int \mathrm{d} \Phi\left(p_1, p_2\right) \phi_b\left(p_1, p_2\right)\left|p_1, \xi_1 ; p_2, \xi_2\right\rangle \,,
	\]
	where $a_1$ and $a_2$ are spin (little group) indices and the $\xi_i$ are vectors in the little group space which pick a specific orientation for the spin expectation value. 
    We evolve this state with the $S$-matrix in the usual way to give a final state
	\[
	S\ket \Psi  = \ket{\Psi} +  \int & \sd \Phi\left(p_1',p_2'\right) \dd^4 q \, \phi_b\left(p_{1}'-q, p_{2}'+q\right) \ket{p_{1}', a_{1}' ; p_{2}', a_{2}'}\\& \times \hat{\delta}\left(2p_1'\cdot q - q^2\right)\hat{\delta}\left(2p_2'\cdot q + q^2\right) i\mathcal{A}_4(p_1',p_2',q)_{a_{1} a_{2}}^{a'_{1} a'_{2}} \, \xi_1^{a_{1}} \xi_2^{a_{2}} \,.
	\]
	
	We wish to write this final state in terms of the eikonal given in eq. \eqref{eikonal_spin1}, and we can do so by noting that the shifted delta functions in the state above can be written in terms of the \emph{unshifted} functions via a translation operator, i.e.
	\[
	\hat{\delta}(2p_1'\cdot q - q^2)\hat{\delta}(2p_2'\cdot q + q^2) = e^{q\cdot Y}\hat{\delta}(2p_1'\cdot q)\hat{\delta}(2p_2'\cdot q)e^{-q\cdot Y},
	\]
	where we have defined
	\[
	Y_\mu = \frac12\frac{\pd}{\pd p_2'^\mu}-\frac12\frac{\pd}{\pd p_1'^\mu}.
	\]
	Adopting the notation $\widetilde{X} = e^{q\cdot Y}Xe^{-q\cdot Y}$, and suppressing the little group indices for the moment, we can include an amplitude to find
	\[
    \label{eq:commutators}
	\hat{\delta}(2p_1'\cdot q - q^2)\hat{\delta}(2p_2'\cdot q + q^2)\cl{A}_4 &= e^{q\cdot Y}\hat{\delta}(2p_1'\cdot q)\hat{\delta}(2p_2'\cdot q)e^{-q\cdot Y}\cl{A}_4\\
	&= \hat{\delta}(2\tilde{p}_1\cdot q)\hat{\delta}(2\tilde{p}_2\cdot q)\left[\tilde{\cl{A}}_4 + e^{q\cdot Y}[e^{-q\cdot Y},\cl{A}_4]\right] \,.
	\]
    In these expressions, the differential operator $e^{\pm q \cdot Y}$ acts on all quantities to its right.
    Notice that we have rewritten the four-point amplitude $\cl{A}_4$ in terms of a translated amplitude $\tilde{\cl{A}}_4$ at the expense of picking up a commutator term.
    
	We recognise the first term on the right of equation~\eqref{eq:commutators} as the inverse Fourier transform of the eikonal in eq. \eqref{eikonal_spin1}, i.e.
	\[
	i\del(2p_1\cdot q)&\del(2p_2\cdot q) \amp_4(s,q^2)^{a_1' a_2'}_{a_1,a_2} \\&= \frac{1}{\hbar^4}\int \d^4 x \, e^{iq\cdot x/\hbar}\left\{\exp\bigg({i\chi(\Pi(p_1,p_2)\cdot x,p_1,p_2)/\hbar}\bigg)^{a_1',a_2'}_{a_1,a_2} - \delta^{a_1'}_{a_1}\delta^{a_2'}_{a_2}\right\} \,,
	\]
	and so we conclude that the final state can be written
	\[
	S\ket \Psi  = &\int  \sd \Phi\left(p_1',p_2'\right) \dd^4 q \, \phi_b\left(p_{1}'-q, p_{2}'+q\right) \ket{p_{1}'; p_{2}'}\\& \times 
	\left(e^{q\cdot Y}\int \sd ^4x e^{iq\cdot x}e^{i\chi(\Pi(p_1',p_2')\cdot x,p_1',p_2')}e^{-q\cdot Y} + e^{q\cdot Y}\hat{\delta}(2p_1'\cdot q)\hat{\delta}(2p_2'\cdot q)[e^{-q\cdot Y},\cl{A}_4]\right), \,
	\]
	or, restoring the little group indices and expressing everything in terms of shifted functions,
	\[
	S\ket \Psi  = &\int \sd \Phi\left(p_1',p_2'\right) \dd^4 q \, \phi_b\left(p_{1}'-q, p_{2}'+q\right) \ket{p_1',a_1';p_2',a_2'}\\& \times 
	\left(\int \sd ^4x e^{iq\cdot x}\exp\bigg({i\chi(x_\perp,\tilde{p}_1,\tilde{p}_2)}\bigg)^{a_1',a_2'}_{a_1,a_2} + \hat{\delta}(2\tilde{p}_1\cdot q)\hat{\delta}(2\tilde{p}_2\cdot q)\left[\cl{A}_4 - \tilde{\cl{A}}_4\right]^{a_1',a_2'}_{a_1,a_2}\right)\xi^{a_1}\xi^{a_2}. \,
	\]
	For the final state to be solely described by the eikonal, it's essential that the second term becomes negligible in the classical limit. This is achieved if the amplitudes are shift-invariant up to $\cl{O}(\hbar)$ corrections --- so we expect that the amplitudes obey the property
	\[
	\hat{\delta}(2\tilde{p}_1\cdot q)\hat{\delta}(2\tilde{p}_2\cdot q)e^{-q\cdot Y}[e^{q\cdot Y},\cl{A}(q)] = \cl{O}(\hbar).
	\]
	One can easily check this property for the usual Mandelstam $s$ and $t$ invariants: $Y^\mu s^n = Y^\mu t^n = 0$, which implies that an amplitude which is a pure polynomial in $s,t$ is unaffected by the shift. However, the mass term is an exception, since for example
	\[
	\tilde{m}_1^2 = m_1^2 + \tilde{p}_1\cdot q + \cl{O}(\hbar). 
	\]
	Thankfully, this additional term is taken care of by the delta functions and we conclude that the amplitude with external scalars is indeed invariant at all orders. Including massive, spinning external particles complicates the situation, introducing terms of the form $\epsilon\cdot p$ and $\epsilon\cdot\epsilon$, where $\epsilon$ is a polarization tensor (or spinor) for the massive spin-$s$ particle in the initial or final state.
	
For the moment, we will confine ourselves to studying linear in spin amplitudes at one loop order, which means we need to show that $e^{-q\cdot Y}[e^{q\cdot Y},\cl{A}(q)] = \cl{O}(\hbar)$ at $\cl{O}(s_i)$. In the classical limit, we can exchange all polarizations for spin-vectors $s_i^\mu$, and so we conclude that the scattering amplitude is a Lorentz scalar function of the vectors $\{s_1,p_1,p_2,q\}$. Since we have established that Mandelstam variables and mass terms are shift-invariant (up to terms killed by the delta function), we only need to consider the possible terms that arise involving spin. At linear order in spin, using the vectors at our disposal, we can only construct the scalar $\varepsilon(p_1,p_2,q,s_1)$. This is clearly invariant under the explicit shifts of the momentum $p_i\rightarrow p_i \pm q/2$, however we also need to check that it is invariant under the shift of the spin. Since $q$ is of order $\hbar$, we can express the shift as a boost
	\[
	s_1(p_1+\frac12 q) = s_1(p_1) + \frac{1}{2m_1^2}p_1^\mu(q\cdot s_1),
	\]
	such that the spin contribution also vanishes. Higher orders terms in this expansion are quantum corrections.
	
	We have shown that $\varepsilon(p_1,p_2,q,s_1)$ is invariant under the shift, and therefore so is the amplitude: $\tilde{\cl{A}}_4 - \cl{A}_4 = 0$.
	
	With this in mind, let's consider the change in an observable $\bb{O}$ given by
	\[
	\Delta \bb{O}=\braket{\Psi|S^{\dagger} \hat{\bb{O}} S|\Psi}-\braket{\Psi|\hat{\bb{O}}| \Psi}=\braket{\Psi|S^{\dagger}[\hat{\bb{O}}, S]|\Psi} \,.
	\]
	In the limit of large classical spin and converting all commutators into Poisson brackets, we can express this final state in terms of purely scalar functions as \cite{Cristofoli:2021jas}
	\[
	[\hat{\bb{O}}, S]\ket{\Psi}  &=\frac{1}{\hbar^4} \int \mathrm{d} \Phi\left(p_1^{\prime}, p_2^{\prime}\right) \int \hat{\mathrm{d}}^4 q \mathrm{~d}^4 x \phi_b\left(p_1^{\prime}-q, p_2^{\prime}+q\right) e^{i q \cdot x / \hbar} \\
	& \times\left(\left\{\bb{O}_1\left(p_1^{\prime}\right), e^{i \chi\left(x_{\perp}, \tilde{p}_1,\tilde{p}_2 ;\langle \tilde{w}\rangle\right) / \hbar}\right\}_{\text {B.H. }}+\bb{O}_1(q) e^{i \chi\left(x_{\perp}, \tilde{p}_1,\tilde{p}_2 ;\langle \tilde{w}\rangle\right) / \hbar}\right)\left|p_1^{\prime}, \xi_1 ; p_2^{\prime}, \xi_2\right\rangle,
	\]
	where we have adopted the notation
	\[
	\left\{\bb{O}_1\left(p_1^{\prime}\right), e^{i \chi\left(x_{\perp}, s ;\langle \tilde{w}\rangle\right) / \hbar}\right\}_{\text {B.H. }} \equiv e^{i \chi / \hbar}\left(-\{\bb{O}, \chi\}-\frac{1}{2}\{\chi,\{\bb{O}, \chi\}\}+\cdots\right).
	\]
	Since all our functions are now scalar (the expectation values of the operators), we can perform the $q,x$ integrals by stationary phase, finding the following conditions on $q$ and $x$
	\[\label{spcondQ}
	Q^\mu &= -\frac{\pd}{\pd x_\mu}\chi(\Pi(p_1' + \frac12 Q,p_2'-\frac12 Q)\cdot x,p_1'+\frac12 Q,p_2'-\frac12 Q, \tilde{s}_1,\tilde{s}_2)\\
	& = -\frac{\pd}{\pd x_\mu}\chi(x_\perp,\tilde{p}_1,\tilde{p}_2,\tilde{s}_1,\tilde{s}_2),
	\]
	and
	\[\label{spcondX}
	X^\mu = b^\mu - \frac{\pd}{\pd Q_\mu}\chi(x_\perp,\tilde{p}_1,\tilde{p}_2,\tilde{s}_1,\tilde{s}_2).
	\]
	Notice that \textit{all} functions that appear in the eikonal that are dependent on $p_1',p_2'$ must be evaluated on the shifts determined by the translation.
	
	The final state can then be expressed as
	\[
	[\hat{\bb{O}}, S]\ket{\Psi}  &=\frac{1}{\hbar^4} \int \mathrm{d} \Phi_b\left(p_1^{\prime}, p_2^{\prime}\right) \phi\left(p_1^{\prime}-Q, p_2^{\prime}+Q\right) e^{i Q \cdot X / \hbar} \\
	& \times\left(\left\{\bb{O}_1\left(p_1^{\prime}\right), e^{i \chi\left(x_{\perp}, \tilde{p}_1,\tilde{p}_2 ;\langle \tilde{w}\rangle\right) / \hbar}\right\}_{\text {B.H. }}+\bb{O}_1(Q) e^{i \chi\left(x_{\perp}, \tilde{p}_1,\tilde{p}_2 ;\langle \tilde{w}\rangle\right) / \hbar}\right)\left|p_1^{\prime}, \xi_1 ; p_2^{\prime}, \xi_2\right\rangle,
	\]
	where now we note that
	\[
	\tilde{p}_1 = p_1' - \frac12 Q = p_1 + \frac12 Q,~~~~~\tilde{p}_2 = p_2' + \frac12 Q = p_2 - \frac12 Q.
	\]
	We can of course evaluate $\bra{\Psi}S^\dagger$ by stationary phase too, finding
	\[
	\bra{\Psi}S^\dagger = \int d\Phi(p_1,p_2)\phi_b(p_1-Q,p_2+Q)e^{-iQ\cdot X/\hbar}e^{-i \chi\left(x_{\perp}, \tilde{p}_1,\tilde{p}_2 ;\langle \tilde{w}\rangle\right) / \hbar}\bra{p_1,\xi_1;p_2,\xi_2},
	\]
	such that
	\[
	\Delta\bb{O}_1 &= \braket{\Psi|S^\dagger[\bb{O}_1,S]\Psi} \\
	&=\Lexp e^{-i \chi\left(x_{\perp}, \tilde{p}_1,\tilde{p}_2 ;\langle \tilde{w}\rangle\right) / \hbar}\left\{\bb{O}_1(p_1),e^{i \chi\left(x_{\perp}, \tilde{p}_1,\tilde{p}_2 ;\langle \tilde{w}\rangle\right) / \hbar}\right\}_{B.H.} + \bb{O}_1(Q)\Rexp\\
	&= \bb{O}_1(Q) - \{\bb{O}_1(p_1),\chi\} - \frac12\{\chi,\{\bb{O}_1(p_1),\chi\}\} + \cdots \,.
	\]
    We emphasise that after stationary phase, tilde variables (such as $\tilde p_1$ etc) involve the classical shift $Q$ rather than the quantum shift $q$.
	If we consider the impulse, taking $\bb{O}_1 = \bb{P}^\mu_1$, then only the first term is non-zero and we find the expected result
	\[
	\Delta p_1^\mu = Q^\mu = -\frac{\pd}{\pd x_\mu}\chi(x_\perp,\tilde{p}_1,\tilde{p}_2).
	\]
	The \textit{angular} impulse, or spin kick, is determined by choosing $\bb{O}_1 = \bb{W}_1^\mu$, which now does give rise to an expansions in Poisson brackets involving the spin vector given by
	\[
	s_{ij}^\mu(p)\hat{\delta}(p-p') = \frac{1}{m}\braket{p,a_i|\bb{W}^\mu|p',a_j},
	\]
	such that
	\[
	\Delta s_1^\mu=s_1^\mu\left(Q\right)-\left\{s_1^\mu, \chi\right\}-\frac{1}{2}\left\{\chi,\left\{s_1^\mu, \chi\right\}\right\}+\cdots,
	\]
 where the spin vector without indices is understood as being an expectation value.
We can think of this expression for the spin-kick in terms of a boost and a little group rotation, covariantly expressed as
\[
W^{\mu}_{~\nu}(p,n) = \exp\left(\frac{\theta}{m}\varepsilon^\mu_{~\nu}(p,n)\right).
\]

To see that this generates the terms above, we note that, at leading order, we have
\[
\{s_1^\mu,\chi\} = \frac{1}{m_1}\varepsilon^{\mu}\left(\frac{\pd\chi}{\pd s_1},p_1,s_1\right),~~~~~\frac12\{\chi,\{s_1^\mu,\chi\}\} = \frac12 \frac{1}{m_1^2}\varepsilon^{\mu\nu}\left(\frac{\pd\chi}{\pd s_1},p_1\right)\varepsilon_{\nu}\left(\frac{\pd\chi}{\pd s_1},p_1,s_1\right),
\]
and similarly at higher order in $\chi$. This tells us that the infinite sum of Poisson brackets is nothing more than a little group rotation around the axis $n^\mu = \frac{1}{\theta}\frac{\pd\chi}{\pd s_\mu}$, i.e.
\[
s_1^\mu -\left\{s_1^\mu, \chi\right\}-\frac{1}{2}\left\{\chi,\left\{s_1^\mu, \chi\right\}\right\} + \cdots = W^\mu_{~\nu}\left(p_1,\frac{\pd\chi}{\pd s_1}\right)s_1^\nu 
\]

We can represent the final spin as the composition of a boost and rotation acting on the initial spin
	\[\label{spinLorentz}
	s_1^\mu(p_1) + \Delta s_1^\mu &= W^{\mu}_{~\nu}K^\nu_{~\rho}s_1^\rho(p_1) \\
	&= \exp\left(-\frac{1}{m_1}\varepsilon^{\mu}_{~\nu}\left(\frac{\pd\chi}{\pd s},p_1\right)\right)\exp\left(\frac{2}{m_1^2}Q^{[\nu} p_{1\rho]}\right)s_1^\rho(p_1)\\
	&= s_1^\mu - (s_1(p_1)\cdot Q)p_1^\mu - \left\{s_1^\mu, \chi\right\}-\frac{1}{2}\left\{\chi,\left\{s_1^\mu, \chi\right\}\right\}+\cdots
	\]
 
 where on the last line we have expressed the ``direct'' term at first order in $Q$ as
	\[\label{direct}
	s_1^\mu(Q) = \frac{1}{m_1^2}\left((Q\cdot p_1)s_1^\mu(p_1) - (s_1(p_1)\cdot Q)p_1^\mu\right),
	\]
 and dropped the higher order in $Q$ terms. It is useful to point out that the condition $s_1\cdot p_1 = (s_1 + \Delta s_1)(p_1 + \Delta p) = 0$ coupled with Lorentz invariance requires that $p_1^\mu + \Delta p_1^\mu$ transform in the same way.
	\section{Classical spin dynamics from the Eikonal}
 \label{sect:Observables}
	To compute either the impulse or the spin kick above linear order in $G$, we need to evaluate the eikonal on the shifted variables --- we shift $p_i'\rightarrow p_i' \pm \frac12 \Delta p$ along with $s_i'\rightarrow s_i' \pm \frac12 \Delta s$, and we will need the value of $x_\perp^\mu$ at the appropriate order in $G$. Considering only particle one to be spinning, we can expand the stationary phase condition in eq. \eqref{spcondX} as
	\[
	X^\mu &= b^\mu - \frac{\pd\chi}{\pd x_\perp^\nu}\frac{\pd x_\perp^\nu}{\pd Q_\mu} - \frac{\pd\tilde{\chi}}{\pd Q_\mu}\bigg|_{Q=0}\\
	&= b^\mu + Q_\nu\frac{\pd x_\perp^\nu}{\pd Q_\mu} - \frac{\pd\tilde{\chi}}{\pd Q_\mu}\bigg|_{Q=0}
	\] 
	where the second term reflects the implicit $Q$ dependence on the unshifted eikonal (where all of the  dependence is in $x_\perp$) and the third the explicit dependence caused by the shifts. Notice that for amplitudes involving only scalars, the second term is absent due to the invariance of the Mandelstam variables under the shift.
	
	To find $x_\perp$, we follow the strategy in appendix B of \cite{Cristofoli:2021jas} and write
	\[
	x_\perp^\mu = X^\mu - (X\cdot e_0)e_0^\mu + (X\cdot e_q)e_q^\mu,
	\]
	where $x_\perp\cdot e_0 = x_\perp\cdot e_q = 0$ and 
	\[
	\left\{\begin{array}{l}
		e_0^\mu \equiv N_0\left(\tilde{p}_1^\mu+\tilde{p}_2^\mu\right)\\
		e_q^\mu \equiv N_q\left(\tilde{p}_1^\mu-\tilde{p}_2^\mu\right)-N_{0 q}\left(\tilde{p}_1^\mu+\tilde{p}_2^\mu\right)
	\end{array}\right.
	\]
	Let's concern ourselves with finding $x_\perp$ to linear order in both $G$ and spin $s_1^\mu$. At this order, the classical dependence\footnote{As noted earlier, the Mandelstam variables are \textit{not} explicitly dependent on $\tilde{p}_i$ since they are shift invariant.} on spin and $\tilde{p}_i$ is always of the form $\varepsilon(x_\perp,\tilde{p}_1,\tilde{p}_2,\tilde{s}_1)$, whose coefficient doesn't explicitly depend on $\tilde{p}_1$, $\tilde{p}_2$ or $\tilde{s}_1$. Ultimately, we wish to express observables in terms of initial and final spins and momentum $s_i$ and $p_i$, so we will need to relate these to their shifted counterparts $\tilde{s}_i$ and $\tilde{p}_i$, as these are what will show up in the eikonal. Lorentz invariance and the spin orthogonality condition  $s\cdot p = (s + \Delta s)\cdot(p + \Delta p) = 0$ ensure that both $p_i^\mu$ and $s_i^\mu$ must undergo the same Lorentz transformation relating them to their shifted counterparts, i.e.
	\[
	\tilde{s}_1^\mu(\tilde{p}_1) = \Lambda\left(\frac12 Q, p_1'\right)^\mu_{~\nu}s_1^\nu(p_1'),~~~~~\tilde{p}_1^\mu = \Lambda\left(\frac12 Q, p_1'\right)^\mu_{~\nu}p_1'^\nu.
	\]
	Since by eq. \eqref{spinLorentz} we know how the spin-vector transforms during the course of a collision, we must assume that after the shift we have performed half of this transformation, and thus the correct Lorentz transformation is of the form
	\[
	\Lambda\left(\frac12 Q, p_1'\right)^\mu_{~\nu} = \exp\left(-\frac{1}{2m_1}\varepsilon^{\mu}_{~\rho}\left(\frac{\pd\chi}{\pd s},p_1\right)\right)\exp\left(\frac{1}{m_1^2}Q^{[\rho} p_{1\nu]}\right),
	\]
	which gives
	\[
	\tilde{s}_1^\mu(\tilde{p}_1) = s_1^\mu(p_1) + \frac12 \Delta s_1^\mu,~~~~~~\tilde{p}_1^\mu = p_1^\mu + \frac12 \Delta p^\mu.
	\]
	\subsection{Shifted Eikonal Functions}
	In order to compute observables at order $G^2$, we will need the tree-level and one-loop eikonal functions evaluated on the shifted momentum $\tilde{p}_i$ and shifted spin $\tilde{s}_i$. We will restrict our attention to the linear in spin, $\cl{O}(G^2)$ sector for now, and so we will consider the spinor-scalar case, where the tree-level shifted phase is given by
	\[
	\chi_{1~b}^a &= \int \hat{d}^4q~\hat{\delta}(2p_1'\cdot q)\hat{\delta}(2p_2'\cdot q)e^{-iq\cdot x_\perp}\cl{M}^{tree}_4[1,1',2,2']^a_{~b}\Bigg|_{p_1'\rightarrow \tilde{p}_1,~~~p_2'\rightarrow \tilde{p}_2}.
	\]
	The classical contributions to the $t$-channel unshifted spinor-scalar amplitude, in terms of $p_1'$ and $p_2'$, is given by
	\[
	\cl{M}_{4~b}^a &=  -\frac{\kappa^2}{16q^2}\bigg[\frac{\alpha_1^{(0)}}{m_1}\bar{u}^a(p_1')u_b(p_1'-q) - \frac{2\alpha_1^{(1,1)}}{m_1^2m_2}\varepsilon_{\mu\nu\rho\sigma}q^\nu p_2'^\mu p_1'^\rho s^\sigma(p_1')^a_{~b}\bigg]
	\]
	where $a,b$ are little group indices and we have ignored terms set to zero by the delta functions in the eikonal. 
	The coefficients in the case of gravity are given by 
	\[
	\alpha_1^{(0)} =
	4m_1^2m_2^2(2\gamma^2-1),~~~~~\alpha_1^{(1,1)} =
	-8im_1^2m_2^2\gamma.
	\]	
	At this point, it is tempting to express $\bar{u}^a(p_1')u_b(p_1'-q) = 2m_1\delta^a_b + \cl{O}(\hbar)$, however this would be premature: we need to make sure we have found all possible spin structures before taking the $\hbar\rightarrow 0$ limit. We will perform this step \textit{after} evaluating the amplitude on the shifted momenta $\tilde{p}_i$, taking $p_i'\rightarrow \tilde{p}_i$. We find
	\[
	\cl{M}_{4~b}^a &=  -\frac{\kappa^2}{16q^2}\bigg[\frac{\alpha_1^{(0)}}{m_1}\bar{u}^a(\tilde{p}_1)u_b(\tilde{p}_1-q) - \frac{2\alpha_1^{(1,1)}}{m_1^2m_2}\varepsilon_{\mu\nu\rho\sigma}q^\nu \tilde{p}_2^\mu \tilde{p}_1^\rho s^\sigma(\tilde{p}_1)^a_{~b}\bigg]
	\]
	In the perturbative regime where $p_i'\gg Q$, we can write the remaining spinor contraction as
	\[
	\bar{u}^a(\tilde{p}_1)u_b(\tilde{p}_1-q) &= \bar{u}^a(p_1'-\frac12 Q)e^{\frac{i}{2}\omega_{\mu\nu}S^{\mu\nu}}u_b(p_1'-\frac12 Q)\\
	&= \bar{u}^a(p_1')e^{-\frac{i}{2}\bar{\omega}_{\alpha\beta}S^{\alpha\beta}}e^{\frac{i}{2}\omega_{\mu\nu}S^{\mu\nu}}e^{\frac{i}{2}\bar{\omega}_{\rho\sigma}S^{\rho\sigma}}u_b(p_1'),
	\]
	where\footnote{We take $\tilde{p}_1^2 = m_1^2 + \frac{Q^2}{4} \simeq m_1^2$ here.}
	\[
	\omega_{\mu\nu} = -\frac{1}{m_1^2}(\tilde{p}_{1\mu}q_\nu - \tilde{p}_{1\nu}q_\mu),~~~~~\bar{\omega}_{\mu\nu} = -\frac{1}{2m_1^2}(p'_{1\mu}Q_\nu - p'_{1\nu}Q_\mu).
	\]
	We can use the BCH formula to express the exponentials to first order as
	\[
	e^{-\frac{i}{2}\bar{\omega}_{\alpha\beta}S^{\alpha\beta}}e^{\frac{i}{2}\omega_{\mu\nu}S^{\mu\nu}}e^{\frac{i}{2}\bar{\omega}_{\rho\sigma}S^{\rho\sigma}} &= e^{\frac{i}{2}\omega_{\mu\nu}S^{\mu\nu}} + \frac{i}{2}\bar{\omega}_{\alpha\beta}[e^{\frac{i}{2}\omega_{\mu\nu}S^{\mu\nu}},S^{\alpha\beta}] + \cl{O}(\bar{\omega}^2)\\
	&= e^{\frac{i}{2}\omega_{\mu\nu}S^{\mu\nu}} - \frac{1}{4}\bar{\omega}_{\mu\nu}\omega_{\alpha\beta}[S^{\mu\nu},S^{\alpha\beta}] + \cl{O}(\bar{\omega}^2). 
	\]
	Using the Lorentz algebra, along with $p_{1\mu}'\bar{u}^a(p_1')S^{\mu\nu}u_b(p_1') = 0$, we find
	\[
	\bar{u}^a(\tilde{p}_1)u_b(\tilde{p}_1-q) &= 2m_1\delta^a_{~b} + \frac{i}{2m_1^2}Q_\mu q_\nu \bar{u}^a(p_1')S^{\mu\nu}u_b(p_1')\\
	&= 2m_1\delta^a_{~b} + \frac{i}{m_1^2}\varepsilon(Q,q,p_1',s^a_{1~b}(p_1')),
	\]
	where we have used
	\[
	S^{\mu\nu a}_{~~~~b} = \frac{1}{2m}\bar{u}^a(p)S^{\mu\nu}u_b(p) =  \frac{1}{m}\varepsilon^{\mu\nu}(p,s^a_{~b}).
	\]
	Plugging this in then gives the tree-level amplitude shifted to $\cl{O}(Q,\hbar)$ as required
	\[
	\tilde{\cl{M}}_{4~b}^a &=  -\frac{\kappa^2}{8q^2}\bigg[\alpha_1^{(0)}\left(\delta^a_{~b} + \frac{i}{m_1^3}\varepsilon\left(Q,q,p_1',s^a_{1~b}(p_1')\right)\right) - \frac{\alpha_1^{(1,1)}}{m_1^2m_2}\varepsilon(q, \tilde{p}_1, \tilde{p}_2, s(\tilde{p}_1)^a_{~b})\bigg].
	\]
	With this in hand, we can compute the second component of $X^\mu$, for which we will need
	\[
	\frac{\pd\tilde{\chi}_1}{\pd Q_\mu}\Bigg|_{Q=0} &= \int \hat{d}^4q~\hat{\delta}(2\tilde{p}_1\cdot q)\hat{\delta}(2\tilde{p}_2\cdot q)e^{-iq\cdot x_\perp}\frac{\pd}{\pd Q_\mu}\tilde{\cl{M}}_4\Bigg|_{Q=0}\\
	&= \frac{G}{4m_1^3m_2^2\sqrt{\gamma^2-1}b^2}\left(m_2\alpha_1^{(0)}\varepsilon^\mu(s_1,u_1,b) + i\alpha_1^{1,1}\left(m_1\varepsilon^\mu(s_1,u_1,b) + m_2\varepsilon^\mu(s_1,u_2,b)\right)\right),
	\]
	where we have used
	\[
	\varepsilon(q,\tilde{p}_1,\tilde{p}_2,s_1(\tilde{p}_1)) =  \varepsilon(q,\tilde{p}_1,\tilde{p}_2,s_1(p_1') + \frac12\Delta s_1) = \varepsilon(q,\tilde{p}_1,\tilde{p}_2,s_1(p_1')) + \cl{O}(G).
	\]
In order to now compute $x_\perp^\mu$, we project to the plane perpendicular to $\tilde{p}_1$ and $\tilde{p}_2$ via
	\[
	x_\perp^\mu = \Pi^\mu_{~\nu}(\tilde{p}_1,\tilde{p}_2)X^\nu = X^\mu - (X\cdot e_0)e_0^\mu + (X\cdot e_q)e_q^\mu.
	\]
    It is particularly useful to work in a basis
    \[
    \left\{b^\mu, u_1^\mu, u_2^\mu, \varepsilon^\mu(b,u_1,u_2)\right\} = \left\{b^\mu, u_1^\mu, u_2^\mu, L^\mu\right\},
    \]
    where $L^\mu = \varepsilon^{\mu}(b,u_1,u_2)$ is related to the angular momentum\footnote{Note, however, that this doesn't have dimensions of angular momentum. It differs by a normalisation factor from, for example, Ref. \cite{Jakobsen:2022zsx}}.
    This basis is particularly useful when considering contributions to observables that are projected into the $\tilde{p}_1$-$\tilde{p}_2$ plane, since it allows us to immediately throw away any terms that happen to be in the direction $u_1^\mu$ or $u_2^\mu$ at the maximal order in $G$ that we care about, since the projection will always generated higher order terms. With this in mind, we find that
	\[
	\Pi^{\mu}_{~\nu}\frac{\pd\tilde{\chi}_1}{\pd Q_\nu}\Bigg|_{Q=0} = \frac{G(m_2\alpha_1^{(0)} + i(m_1+\gamma m_2)\alpha_1^{(1,1)})s_1\cdot u_2}{4m_1^3m_2^2(\gamma^2-1)^{3/2}b^2}\varepsilon^\mu(b,u_1,u_2)
	\]
	We can compute the products of $X$ with the projected directions
	\[
	X\cdot e_0 &= b\cdot e_0 = 0,\\
	X\cdot e_q &= b\cdot e_q + \cl{O}(G^2)
	\]
	At order $G$ then, we find that $x_\perp$ is given by
	\[
	x_\perp^\mu &= b^\mu  - (b\cdot e_q)e_q^\mu +\frac{G(m_2\alpha_1^{(0)} + i(m_1+\gamma m_2)\alpha_1^{(1,1)})s_1\cdot u_2}{4m_1^3m_2^2(\gamma^2-1)^{3/2}b^2}\varepsilon^\mu(b,u_1,u_2)\\
	&=  b^\mu - \frac{b\cdot \Delta p}{2m_1m_2(\gamma^2-1
		)}K^\mu + \frac{G(m_2\alpha_1^{(0)} + i(m_1+\gamma m_2)\alpha_1^{(1,1)})s_1\cdot u_2}{4m_1^3m_2^2(\gamma^2-1)^{3/2}b^2}\varepsilon^\mu(b,u_1,u_2)
	\]
	where we have defined
	\[
	K^\mu = (m_1+\gamma m_2)u_2^\mu - (m_2+\gamma m_1)u_1^\mu.
	\]
    It may be helpful to note that this Lorentz vector, expressed in the centre-of-mass frame, is proportional to the spatial momentum of one of the particles.
 
	At the leading order, we can write
\[
b\cdot\Delta p &= b_\mu\frac{\pd\tilde{\chi}_1}{\pd x_\mu}\\
&= \frac{G(m_1b^2\alpha_1^{(0)}+i\alpha_1^{(1,1)}L\cdot s_1)}{2m_1^2m_2\sqrt{\gamma^2-1}b^2}
\]	
	and therefore that the leading order contribution to $x_\perp$ is
	\[\label{xperpG}
	x_\perp^\mu &= b^\mu + \frac{G(m_1b^2\alpha_1^{(0)}+i\alpha_1^{(1,1)}L\cdot s_1)}{2m_1^2m_2\sqrt{\gamma^2-1}b^2}
	K^\mu + \frac{G(m_2\alpha_1^{(0)} + i(m_1+\gamma m_2)\alpha_1^{(1,1)})s_1\cdot u_2}{4m_1^3m_2^2(\gamma^2-1)^{3/2}b^2}L^\mu.
	\]
	We have now computed all the ingredients we need to compute observables up to order $\cl{O}(G^2)$, but before we do it's useful to summarise the overall process.
\subsection{Computing Spinning Observables from the Eikonal}
The impulse is obtained from the Eikonal phase using the following formula
	\begin{align}
		\D p^{\mu}&=-\Pi^\mu_{~\nu}\frac{\pd \tilde{\chi}}{\pd x_{\perp\nu}},
		\label{covLinearImpulse}
\end{align}
where the tilde reminds us that we need to fully evaluate all of the shifted variables as in the previous subsection. With this in mind, it is useful to work with mass-rescaled variables
 \[
 \tilde{u}_i \equiv \frac{\tilde{p}_i}{\tilde{m}_i},~~~~~\tilde{a}_i \equiv \frac{\tilde{s}_i}{\tilde{m}_i},
 \]
and we recall that the dependence is of the form
	\begin{align}
		\tilde{\chi} = \tilde{\chi} (\tilde{u}_i,\tilde{a}_i,x_\perp),\quad
		\D p = \D \tilde{p} (\tilde{u}_i,\tilde{a}_i,x_\perp),
	\end{align}
As we have explicitly shown, we can express the shifts in terms of the more familiar variables 
	\begin{align}
		\nonumber
		x_\perp^\mu&=b^\mu\,+\frac{\D b^{\mu}}{2},
		\\
		\label{eq:shifted_vars}
		\tilde{u}_i^\mu&=u_i^\mu+\frac{\D u_{i}^{\mu}}{2},
		\\
		\tilde{a}_i^\mu&=a_i^\mu+\frac{\D a_{i}^{\mu}}{2}.
		\nonumber
	\end{align}

Let's motivate this by computing some observables at $\cl{O}(G^2)$, considering the eikonal phases of the form
\begin{align}
    \tilde{\chi}=\tilde{\chi}_1+\tilde{\chi}_2+\mathcal{O}(G^3).
\end{align}
The dependence up to the linear-in-spin order is given by
	\begin{align}
		\label{chi1}
		\tilde{\chi}_1= 
		-\frac{2G}{4m_1m_2\sqrt{\gamma^2-1}} \,
		\, 
		\big(
		&\alpha_1^{(0)}
		\log{|x_\perp|}+
		i\alpha_1^{(1,1)}
		\frac{\varepsilon(\tilde{a}_1,\tilde{u}_1,\tilde{u}_2,x_\perp)}{|x_\perp|^2}
		\big)
		\,,\\
 \label{chi2}
		\tilde{\chi}_2= 
		\frac{\pi G^2}{4m_1m_2\sqrt{\gamma^2-1}} \,
		\, 
		\big(
	&\alpha_2^{(0)}\frac{1}{|x_\perp|}+
		i\alpha_2^{(1,1)}
		\frac{\varepsilon(\tilde{a}_1,\tilde{u}_1,\tilde{u}_2,x_\perp)}{|x_\perp|^3}
		\big)
		\,,
	\end{align}
	where $|x_\perp|=\sqrt{-x_\perp^2}$, and the explicit coefficients in the case of gravity are
 \begin{align}
     \label{alphascalar}
     \alpha_1^{(0)}&=
		4m_1^2m_2^2(2\gamma^2-1)
		\,,\qquad\ 
		\alpha_2^{(0)}=
		3m_1^2m_2^2(m_1+m_2)(5\gamma^2-1)
		\,,\\
  \label{alphaso}
  \alpha_1^{(1,1)}&=
		-8im_1^2m_2^2\gamma
		\,,\qquad\qquad 
		\alpha_2^{(1,1)}=
		-\frac{im_1^2m_2^2(4m_1+3m_2)(5\gamma^2-3)\gamma}{(\gamma^2-1)}
		\,.
 \end{align}
The now-familiar scalar impulse can be expressed as
\[
		\D p_{1,\,\textrm{scalar}}^{\mu}
		&=
		-\frac{G}{4m_1m_2\sqrt{\gamma^2-1}|b|^2}
\Big(
2\alpha_1^{(0)}+
\frac{G\pi}{|b|}\alpha_2^{(0)}
\Big)x_\perp^\mu
\\
&=
		-\frac{G}{4m_1m_2\sqrt{\gamma^2-1}|b|^2}
		\bigg\{
		\Big(
		2\alpha_1^{(0)}+
		\frac{G\pi}{|b|}\alpha_2^{(0)}
		\Big)b^\mu +
		\frac{G}{2m_1^2m_2^2(\gamma^2-1)^{3/2}}
		(\alpha_1^{(0)})^2
		K^\mu
		\bigg\},
\]
where the second term in the last line is obtained as an iteration\footnote{It is actually possible to determine the iteration terms to all orders in spin from known results since they follow from tree amplitudes. This is not so for the full $\Delta p^\mu$, as one would also need all order in spin one-loop information which is not currently available.} by plugging in eq. \eqref{xperpG}. 

For the terms which are linear in the spin $a_1$, the impulse takes the form
	\begin{align}
		\D p_{1,\,a_1}^{\mu}
		=
		&\frac{G}{4m_1m_2\sqrt{\gamma^2-1}|b|^4}
		\bigg\{  
		i\alpha_1^{(1,1)}\Big(
		(a_1\cdot L)b^\mu+2(a_1\cdot b)L^\mu\Big)+
		\frac{G\pi}{|b|}i\alpha_2^{(1,1)}
		\Big((a_1\cdot L)b^\mu+(a_1\cdot b)L^\mu\Big)
		\nonumber
		\\ 
		&-
		\frac{G}{2m_1^2m_2^2(\gamma^2-1)^{3/2}}
		\Big(
		c_{\Delta p,\textrm{iter}}^{(1)}(a_1\cdot u_2)L^\mu+c_{\Delta p,\textrm{iter}}^{(2)}(a_1\cdot L)K^\mu
		\Big)
		\bigg\},
	\end{align}
	with iteration coefficients
	\begin{align}
		\nonumber    c_{\Delta p,\textrm{iter}}^{(1)}=&(m_1+m_2\gamma)(2\alpha_1^{(0)}i\alpha_1^{(1,1)})
		+m_2\big(({\alpha_1^{(0)}})^2+(\gamma^2-1)({i\alpha_1^{(1,1)}})^2\big),\\
  c_{\Delta p,\textrm{iter}}^{(2)}=&-2\alpha_1^{(0)}i\alpha_1^{(1,1)}.
	\end{align}
 Plugging in the values for the coefficients, we find the spinning impulse
 \[
 \Delta p_1^\mu =&~ \frac{2Gm_1m_2}{\sqrt{\gamma^2-1}b^4}\bigg[\left((2\gamma^2-1)b^2 - 2\gamma\varepsilon(b,a_1,u_1,u_2)\right)b^\mu + 2m_2\gamma b\cdot a_1 L^\mu\bigg]\\
	&+ G^2m_1m_2\bigg[\left(\frac{(2\gamma^2-1)(2(2\gamma^2-1)b^2 + 8\gamma\varepsilon(b,a_1,u_1,u_2)}{(\gamma^2-1)^2b^4}\right)K^\mu\\
	&+\left(\frac{\pi \gamma\left(5\gamma^2-3+\right) b \cdot a_1\left(4 m_1+3 m_2\right)}{4\left(-1+\gamma^2\right)^{3 / 2}\left(-b^2\right)^{5 / 2}}-\frac{2 a_1 \cdot u_ 2\left(4 \gamma m_1-8 \gamma^3 m_1+m_2-4 \gamma^2 m_2\right)}{\left(-1+\gamma^2\right)^2 b^4}\right) L^\mu\\
	&+ \left(\frac{\pi\left(3\left(1-6 \gamma^2+5 \gamma^4\right) b^2 \left(m_1+m_2\right)-2 \gamma\left(5\gamma^2-3\right)\left(4 m_1+3 m_2\right) \varepsilon(b, a_1, u_1, u_2)\right)}{4\left(-1+\gamma^2\right)^{3 / 2}\left(-b^2\right)^{5/2}}\right)b^\mu\bigg]
\]
In rescaled variables, the spin kick can be obtained from the expression
\[
	\Delta a_i^\mu=a_i^\mu\left(\Delta u\right)-\left\{a_i^\mu, \tilde{\chi}\right\}-\frac{1}{2}\left\{\tilde{\chi},\left\{a_i^\mu, \tilde{\chi}\right\}\right\}+\cdots,
\]
This is an iterative expression in the sense that each term in the rotation contains a fundamental bracket $\{a_1^\mu,\chi\}$. At order $\cl{O}(G^2)$, we can take advantage of this to write it in terms of shifted variables as
 \begin{align}
		\D \tilde{a}_{i}^{\mu}(\tilde{u}_i,\tilde{a}_i,x_\perp)&=
(\tilde{u}_i\cdot \D u_{i}) \tilde{a}_i^\mu
  -(\tilde{a}_i\cdot \D u_{i})  \tilde{u}_i^\mu-
		\varepsilon^{\mu\beta\gamma\delta}
		\frac{1}{m_i}\frac{\pd \tilde{\chi}_n}{\pd \tilde{a}_i^\beta}
		\tilde{u}_{i\gamma} 
		\tilde{a}_{i\delta} + \cl{O}(G^3) \,.
	\label{covAngularImpulse}
	\end{align}
Note that the ``direct'' term $a^\mu_i(\Delta u)$ has its origin~\cite{Maybee:2019jus} in the momentum factor of the Pauli-Lubanski vector. 
This factor does \emph{not} contribute (at this order) through any spin commutator.
Consequently, we imposed the constraint
\[\label{eq:constraint}
\{ a^\mu_i(\Delta u), \tilde{\chi} \} = 0 \,
\]
in deriving equation~\eqref{covAngularImpulse}.
It would be interesting to explore the origin of this constraint in more detail.

An analogous process of evaluating this formula with the Eikonal phases eqs. (\ref{chi1})-(\ref{chi2}), and translating into unshifted variables results in the spin kick being expressed as 
	\begin{align}
		\D a_1^{\mu}
		=
		\frac{G}{4m_1m_2\sqrt{\gamma^2-1}|b|^2}
		\bigg\{ 
		&
		\Big(
		2i\alpha_1^{(1,1)}+
		\frac{G\pi}{|b|}i\alpha_2^{(1,1)}
		\Big)
		\Big(
		(a_1\cdot u_2)b^\mu-(a_1\cdot b)u_2^\mu
		\Big)
		\\ \nonumber
		&\Big(
		2\gamma i\alpha_1^{(1,1)}+2\alpha_1^{(0)}+
		\frac{G\pi}{|b|}(\gamma i\alpha_2^{(1,1)}+\alpha_2^{(0)})
		\Big)(a_1\cdot b)u_1^\mu\\ \nonumber
		-
		\frac{G}{2m_1^2m_2^2(\gamma^2-1)^{3/2}}
		&\Big(
		c_{\Delta a_1,\textrm{iter}}^{(1)}
		b^{-2}(a_1\cdot b)b^\mu+
		c_{\Delta a_1,\textrm{iter}}^{(2)}
		(a_1\cdot u_2)u_1^\mu+
		c_{\Delta a_1,\textrm{iter}}^{(3)}
		(a_1\cdot u_2)u_2^\mu
		\Big)\bigg\},
	\end{align}
	with iteration coefficients given by
	\begin{align*}
		c_{\Delta a_1,\textrm{iter}}^{(1)} & = 
		m_ 2 (-1 + \gamma^2) \big(-({\alpha_1^{(0)}})^2 + (-1 + \gamma^2) ({i\alpha_1^{(1,1)}})^2\big), \\
		c_{\Delta a_1,\textrm{iter}}^{(2)} & = 
		m_ 2 (-1 + \gamma^2) \big( 
		2 \alpha_1^{(0)}i\alpha_1^{(1,1)} + \gamma ({i\alpha_1^{(1,1)}})^2\big) + (m_ 1 + m_ 2 \gamma) ({\alpha_1^{(0)}})^2, \\
		c_{\Delta a_1,\textrm{iter}}^{(3)} & = m_ 2 (-1 + \gamma^2) \big(-({i\alpha_1^{(1,1)}})^2\big).
	\end{align*}
For the explicit form of the Eikonal obtained by plugging the values eqs. (\ref{alphascalar})-(\ref{alphaso}), our results  match exactly those of Ref. \cite{Liu:2021zxr} after expressing them in the same basis of vectors. 
 
 More generally, we can use eq. (\ref{eq:shifted_vars}) to express result in terms of our preferred past-infinity variables for any observable. This results in
	\begin{align}
		\D \mathbb{O}_{i}^{(1),\mu}=&
		\D \tilde{\mathbb{O}}_{i}^{(1),\mu}(u_i,a_i,b),\\
		\D \mathbb{O}_{i}^{(2),\mu}=&
		\D \tilde{\mathbb{O}}_{i}^{(2),\mu}(u_i,a_i,b)+\D \mathbb{O}^{(2),\mu}_{i,\textrm{iter}}.
	\end{align}
	where the superscript $(i)$ correspond to the order in the post-Minkowskian approximation, and the observable $\mathbb{O}$ stands for either the momentum $p$, as in eq. (\ref{covLinearImpulse}) or the spin vector $a$, as in eq. (\ref{covAngularImpulse}), but they are now evaluated on the un-shifted variables. The iteration term can be expressed up to the second post-Minkowskian order as 
	\begin{equation}
		\D \mathbb{O}^{(2),\mu}_{i,\textrm{iter}}=
		\sum_i
		\frac{\pd \D \mathbb{O}_{i}^{(1),\mu}}{\pd u_i^\nu}
		\frac{\D u_{i}^{(1),\nu}}{2}+
		\sum_i
		\frac{\pd \D \mathbb{O}_{i}^{(1),\mu}}{\pd a_i^\nu}
		\frac{\D a_{i}^{(1),\nu}}{2}+
		\frac{\pd \D \mathbb{O}_{i}^{(1),\mu}}{\pd b^\nu}
		\frac{\D b^{(1),\nu}}{2}.
		\label{2PMIter}
	\end{equation}
Note that since every impulse entering eq. (\ref{2PMIter}) is taken at the first post-Minkowskian approximation, there is no distinction between $\D \mathbb{O}$ and $\D \tilde{\mathbb{O}}$. Also, while $\D u_i^{(1),\mu}$ and $\D a_i^{(1),\mu}$ are obtained by means of the formulas eqs. (\ref{covLinearImpulse}) and (\ref{covAngularImpulse}), the change in impact parameter $\D b^{(1),\mu}$ is obtained from the second stationary phase condition. Using eq. (\ref{2PMIter}), we can write observables at the second post-Minkowskian approximation compactly as,
	\begin{align}
		\D \mathbb{O}^{(i,j)}
		=
		\frac{G}{4m_1m_2\sqrt{\gamma^2-1}|b|^{2(\zeta_{\mathbb{O}}+i+j)}}
		\bigg\{ &
		\sum_k 
		\mathcal{O}_{\Delta \mathbb{O},\chi}^{(i,j,k)}
		\Big(
		c_{\Delta \mathbb{O},\chi_1}^{(i,j,k)}+
		\frac{G\pi}{|b|}c_{\Delta \mathbb{O},\chi_2}^{(i,j,k)}
		\Big)
		\\ \nonumber
		&-
		\frac{G}{2m_1^2m_2^2(\gamma^2-1)^{3/2}}
		\sum_l
		\mathcal{O}_{\Delta \mathbb{O},\textrm{iter}}^{(i,j,l)}
		c_{\Delta \mathbb{O},\textrm{iter}}^{(i,j,l)}
		\bigg\},
	\end{align}
	where $\mathbb{O}$ means either $p$ or $a_1$ for the impulse or the spin-kick, respectively, and the observable-dependent factor $\zeta_{\mathbb{O}}$ in the denominator is defined by
	\begin{align}
		\zeta_{p}=1,
		\qquad\qquad 
		\zeta_{a_1}=0.
	\end{align}
The indices $i$ and $j$ label the powers of $a_1$ and $a_2$, respectively. The structures $\mathcal{O}_{\Delta p,\chi}$ correspond to contributions which are linear in the Eikonal phase at either the first or second post-Minkowskian order, and so the coefficients $c_{\Delta p,\chi_i}$ depend on one $\alpha_i$, while the $\mathcal{O}_{\Delta p,\textrm{iter}}$ correspond to iterations, and their coefficients $c_{\Delta p,\textrm{iter}}$ depend on products of two $\alpha_1$'s.
We have obtained results up to quadratic order in the spin of the particles, which are presented in an appendix.

\section{Relation to other approaches}
\label{sect:Comparisons}
By application of this prescription, we have derived covariant observables from the eikonal phase, namely the classical impulse and the spin kick, up to the second post-Minkowskian approximation, and up to quadratic order in the spin of the particles. Results at these order were first obtained in covariant form in Ref. \cite{Liu:2021zxr}, by means of the spinning generalisation of the (worldline-based) post-Minkowskian effective field theory (PMEFT) of Ref. \cite{Kalin:2020mvi}. The comparison between both results is straightforward, up to expressing them in a basis of vectors.  We have verified they perfectly agree with each other.

However, there are other existing derivations of these quantities from the Eikonal in the literature, with supposedly different formulations, e.g. those coming from a canonical approach \cite{Bern:2020buy} and those derived from worldline techniques \cite{Jakobsen:2021zvh}. It is interesting to compare these approaches, and so we do so in this section.
 
\subsection{Relation to covariant (worldline) methods}
Let us comment on the relation to another derivation of these observables from an eikonal phase, which was obtained in Ref. \cite{Jakobsen:2021zvh} using a so-called worldline quantum field theory (WQFT). First, we verified that both eikonal phases are a perfect match. Notably, their results for observables are given in terms of the spin tensor, which is related to the (Pauli-Lubanski) spin vector by means of
	\begin{align}
		S^{\mu\nu}=\frac{1}{m}\epsilon^{\mu\nu\sigma\rho}
		p_\sigma S_\rho,\qquad
		S^{\mu}=\frac{1}{2m}\epsilon^{\mu\nu\sigma\rho}
		p_\nu S_{\sigma\rho}.
	\end{align}
	In consequence, we find that to compare their spin-tensor change $\D S^{\mu\nu}$ with our spin kick $\D S^\mu$ we need to evaluate
	\begin{align}
		\D S^{\mu\nu}=\frac{1}{m}\epsilon^{\mu\nu\sigma\rho}
		(\D p_\sigma S_\rho+p_\sigma \D S_\rho+\D p_\sigma \D S_\rho),
	\end{align}
 or alternatively
\begin{align}
		\D S^{\mu}=\frac{1}{2m}\epsilon^{\mu\nu\sigma\rho}
		(\D p_\nu S_{\sigma\rho}+p_\nu \D S_{\sigma\rho}+\D p_\nu \D S_{\sigma\rho}).
	\end{align}
 While our results agree perfectly, it is worth commenting on some similarities and slight differences in our approaches. 
	First, they obtain their result through the relation
 \begin{align}
     \D \tilde{S}^{\mu\nu}=\frac{4}{m}\tilde{S}^{\rho[\mu}\frac{\pd \tilde{\chi}}{\pd \tilde{S}_{\nu]}^{\ \ \rho}},
 \end{align}
 where the tilde once again denotes shifted variables. Indeed, these are naturally chosen by the time-symmetric propagators in their worldline.
 Furthermore, they introduce a shift in their eikonal phase of the form
	\begin{align}
 \label{berlinshiftS}
  \tilde{S}_i^{\mu\nu}	&\rightarrow \tilde{S}_i^{\mu\nu}+2(u_i\cdot S_i)^{[\mu]}u_i^{\nu]}	\\
 \label{berlinshiftb}
  \tilde{b}^\mu &\rightarrow \tilde{b}^\mu + \xi_2 \tilde{u}_2^\mu - \xi_1 \tilde{u}_1^\mu + \tilde{S}_2^{\mu\nu}u_{2,\nu}- \tilde{S}_1^{\mu\nu}u_{1,\nu},
	\end{align}
 where the $\xi_i$ depend in general on $b\cdot u_i$ and $S_i^{\mu\nu}u_{i,\nu}$. While such shifts vanish on-shell, they do produce contributions in the directions of $\tilde{u}_i$ upon taking derivatives of the eikonal phase with respect to $\tilde{b}$ and $\tilde{S}^{\mu\nu}$ in a similar fashion to eq. (\ref{covLinearImpulse}). This shift was demanded to preserve symmetry in their worldline. In order to ensure we rightly reproduce longitudinal terms in the impulse, we also perform the following shift in impact parameter in the Eikonal phase 
	\begin{equation}
		\tilde{b}^\mu \rightarrow \tilde{b}^\mu + \xi_2 \tilde{u}_2^\mu - \xi_1 \tilde{u}_1^\mu,
	\end{equation}
	where 
	\begin{equation}
		\xi_1=\frac{\tilde{b}\cdot (\gamma \tilde{u}_2 - \tilde{u}_1)}{\gamma^2-1},
		\qquad
		\xi_2=\frac{\tilde{b}\cdot (\tilde{u}_2 - \gamma \tilde{u}_1)}{\gamma^2-1}.
	\end{equation}
	However, we have found that the spin shift in eqs. (\ref{berlinshiftS})-(\ref{berlinshiftb}) is unsubstantial for the spin vector, while required for the spin-tensor.
	
	\subsection{Relation to canonical methods (the $D_{SL}$ formula)}
	Concurently with the worldline result of Ref. \cite{Liu:2021zxr}, results at the 2PM and quadratic in spin orders were obtained using scattering amplitudes in Ref. \cite{Kosmopoulos:2021zoq}, building on the formalism introduced in Ref. \cite{Bern:2020buy}.
	The comparison between amplitudes-based results and  worldline approaches is obscured by the fact of them being written in terms of different variables, namely canonical or covariant. Such a comparison is now well understood and fairly explored (see, for example, discussions in Ref. \cite{Jakobsen:2022zsx}).
	
	Instead, we will use our newly-built intuition to provide some insight into the peculiar form of the so-called $D_{SL}$ formulas, which directly obtain impulses and spin kicks from the eikonal phase. These  were reverse-engineered by first obtaining the observables through $\mathcal{O}(G^2)$ using Hamilton's equations. Let us try to shed some light onto their structure. We start with their usual form
	\begin{align}
		\Delta \mathbb{O}^i\, &= -\{ \mathbb{O}^i,\, \chi \}
		-\frac{1}{2}\,\{\chi, \{\mathbb{O}^i,\, \chi\} \}
		-D_{SL}\left(\chi, \{\mathbb{O}^i,\, \chi\} \right)
		+\frac{1}{2}\,\{ \mathbb{O}^i,\,D_{SL}\left( \chi,  \chi \right) \} \,,
		\label{DSL}
	\end{align}
	where $\mathbb{O}$ can stand for either the transverse momentum $p_{\perp}$ or the spin $S$. 
 \iffalse Both relations have been verified to be valid up to $\mathcal{O}(G^2)$ and quadratic order in the spin.\fi 
 Here, we'll show that the iteration piece
 \begin{align}
		\Delta \mathbb{O}_{\textrm{iter}}^{(2),i}=&
		-\frac{1}{2}\,\{\chi_1, \{\mathbb{O}^i, \chi_1\} \}
		-D_{SL}\left(\chi_1, \{ \mathbb{O}^i, \chi_1\} \right)
		+\frac{1}{2}\,\{\mathbb{O}^i,D_{SL}\left( \chi_1,  \chi_1 \right) \} \,.
		\label{Deltapxyiter}
	\end{align}
can be instead written as 
\begin{align}
		\Delta \mathbb{O}_{\textrm{iter}}^{(2),i}=&
		\frac{1}{2}\frac{\pd \D \mathbb{O}^{(1),i}}{\pd S^j}\D S^{(1),j}
		+\frac{1}{2} \frac{\pd \D \mathbb{O}^{(1),i}}{\pd L^j} \D L^{(1),j}
		+\frac{1}{2} \frac{\pd \D \mathbb{O}^{(1),i}}{\pd p^j} \D p^{(1),j}\,,
		\label{Deltapxyiter04}
	\end{align}
 in complete analogy with equation (\ref{2PMIter}), thus replacing the mysterious form of the $D_{SL}$ formula with the more natural evaluation in shifted variables.  
 
To prove this, let us start by reviewing these formulas. The different brackets appearing in eq. (\ref{Deltapxyiter}) are given by
	\begin{align}
		\{p^i_\perp, f \} \equiv -\frac{\pd f}{\pd b^i}  \,, 
		\qquad
		\{S^i, f \}  \equiv \epsilon^{ijk} \, \frac{\partial f}{\partial S^j} S^k \,, 
		\qquad
		D_{SL}\left(f, g \right) \equiv 
		- \,\epsilon^{ijk} S^k\, \frac{\partial f}{\partial S^i}\,\underline{\frac{\partial g}{\partial L^j}}
		\, .
		\label{Brackets}
	\end{align}
	The functions $f$ and $g$ in general depend on $S^i$, $p^i$ and $b^i$ (though $p^i$ is taken as inert in the brackets). For this reason, the $L$ derivative in the $D_{SL}$ bracket is interpreted as 
	\begin{align}
		\underline{ \frac{\pd}{\pd L^j}}\equiv \frac{\pd}{\pd L^j}-\frac{p^j p^k}{p^2}\frac{\pd}{\pd L^k}
		\label{ddLb}
	\end{align}
	Here, we write the formula considering only one spinning particle, but the generalisation to two spinning particles is straightforward.
	We start by writing separately the first two orders in the post-Minkowskian approximation
	\begin{align}
		\label{deltap1pm}
		\Delta \mathbb{O}^{(1),i}   &= -\{ \mathbb{O}^i, \chi_1 \},\\
		\label{deltap2pm}
		\Delta \mathbb{O}^{(2),i}   &= -\{ \mathbb{O}^i, \chi_2 \}+\Delta \mathbb{O}_{\textrm{iter}}^{(2),i}
	\end{align}
	where $\chi_n$ is the nPM Eikonal phase, and we  focus on the iteration term, which is defined in eq. (\ref{Deltapxyiter}).
	Reinserting back eq. (\ref{deltap1pm}) into the iteration, we note we can write
	\begin{align}
		\Delta \mathbb{O}_{\textrm{iter}}^{(2),i}=&
		\frac{1}{2}\,\{\chi_1, \Delta \mathbb{O}^{(1),i}  \}
		+D_{SL}\big(\chi_1, \Delta \mathbb{O}^{(1),i}  \big)
		+\frac{1}{2}\,\{\mathbb{O}^i,D_{SL}\left( \chi_1,  \chi_1 \right) \} \,.
		\label{Deltapxyiter01}
	\end{align}
	Now, from the definition of the Poisson bracket, we know
	\begin{align}
		\{\chi_1, \Delta \mathbb{O}^{(1),i}\} =
		\frac{\pd \Delta \mathbb{O}^{(1),i}}{\pd S^j}
		\{\chi_1, S^j\} = 
		\frac{\pd \Delta \mathbb{O}^{(1),i}}{\pd S^j}\Delta S^{(1),j},
	\end{align}
	where in the last equality we have used eq. (\ref{DSL}) for spin at 1PM
	\begin{align}
		\Delta S^{(1),i}=-\{S^i,\chi_1\}.
	\end{align}
	It is also useful to analyse the expression for the $D_{SL}$ bracket, when one of its arguments is the eikonal phase. This is
	\begin{align}
		D_{SL}(\chi_1,g)&=-\epsilon^{ijk} S^k 
		\frac{\pd \chi_1}{\pd S^i}\frac{\pd g}{\underline{\pd L^j}}\\
		&= \{S^j,\chi_1\} \frac{\pd g}{\underline{\pd L^j}}
		= -\D S^{(1),j} \frac{\pd g}{\underline{\pd L^j}}
		= \D L^{(1),j} \frac{\pd g}{\underline{\pd L^j}},
	\end{align}
	where we have used the conservation of total angular momentum
	\begin{align}
		\D J^i = \D S^i + \D L^i = 0.
	\end{align}
	Then, the iteration can be written as
	\begin{align}
		\Delta \mathbb{O}_{\textrm{iter}}^{(2),i}=&
		\frac{1}{2}\frac{\pd \D \mathbb{O}^{(1),i}}{\pd S^j}\D S^{(1),j}
		+ \frac{\pd \D \mathbb{O}^{(1),i}}{\underline{\pd L^j}} \D L^j
		+\frac{1}{2}\,\{\mathbb{O}^i,\D L^{(1),j} \frac{\pd \chi_1}{\underline{\pd L^j}} \} \,.
	\end{align}
	Now, since the Poisson bracket acts as a differential on its arguments, we find it satisfies the Leibniz-like condition
	\begin{align}
		\{\mathbb{O}^i,\D L^{(1),j} \frac{\pd \chi_1}{\underline{\pd L^j}} \}&=\D L^{(1),j}\{\mathbb{O}^i, \frac{\pd \chi_1}{\underline{\pd L^j}} \}+\{\mathbb{O}^i,\D L^{(1),j}  \}\frac{\pd \chi_1}{\underline{\pd L^j}}\\ 
		&=-\frac{\pd \D \mathbb{O}^{(1),i}}{\underline{\pd L^j}} \D L^{(1),j}+\{\mathbb{O}^i,\D L^{(1),j}  \}\frac{\pd \chi_1}{\underline{\pd L^j}}.
	\end{align}
	where, in the first term of the second line we have used eq. (\ref{deltap1pm}), along with the fact that the action of the Poisson bracket commutes with the derivative ${\pd}/{\underline{\pd L^j}}$. Inserting this back we have
	\begin{align}
		\Delta \mathbb{O}_{\textrm{iter}}^{(2),i}
		=&
		\frac{1}{2}\frac{\pd \D \mathbb{O}^{(1),i}}{\pd S^j}\D S^{(1),j}
		+\frac{1}{2} \frac{\pd \D \mathbb{O}^{(1),i}}{\underline{\pd L^j}} \D L^j
		- \frac{1}{2} \{\mathbb{O}^i,\D L^{(1),j}  \}\frac{\pd \chi_1}{\underline{\pd L^j}}  \,.
		\label{Deltapxyiter02}
	\end{align}
	Recalling eq. (\ref{ddLb}), the iteration becomes
	\begin{align}
		\nonumber
		\Delta \mathbb{O}_{\textrm{iter}}^{(2),i}=&
		\frac{1}{2}\frac{\pd \D \mathbb{O}^{(1),i}}{\pd S^j}\D S^{(1),j}
		+\frac{1}{2} \frac{\pd \D \mathbb{O}^{(1),i}}{\pd L^j} \D L^j\\
		&-\frac{p^jp^k}{2p^2} \frac{\pd \D \mathbb{O}^{(1),i}}{\pd L^k} \D L^j
		- \frac{1}{2} \{\mathbb{O}^i,\D L^{(1),j}  \}\frac{\pd \chi_1}{\underline{\pd L^j}}  \,.
		\label{Deltapxyiter03}
	\end{align}
	Finally, we have verified that up to the orders in spin of our interest, the second line in eq. (\ref{Deltapxyiter03}) satisfies
	\begin{align}
		-\frac{p^jp^k}{2p^2} \frac{\pd \D \mathbb{O}^{(1),i}}{\pd L^k} \D L^j
		- \frac{1}{2} \{\mathbb{O}^i,\D L^{(1),j}  \}\frac{\pd \chi_1}{\underline{\pd L^j}}=
		\frac{1}{2}\frac{\pd \D \mathbb{O}^{(1),i}}{\pd p^j}\D p^{(1),j}.
	\end{align}
	This allows to write the iteration as
	\begin{align*}
		\Delta \mathbb{O}_{\textrm{iter}}^{(2),i}=&
		\frac{1}{2}\frac{\pd \D \mathbb{O}^{(1),i}}{\pd S^j}\D S^{(1),j}
		+\frac{1}{2} \frac{\pd \D \mathbb{O}^{(1),i}}{\pd L^j} \D L^{(1),j}
		+\frac{1}{2} \frac{\pd \D \mathbb{O}^{(1),i}}{\pd p^j} \D p^{(1),j}\,,
	\end{align*}
which had already been presented in eq. (\ref{Deltapxyiter04}). We have thus shown that the iteration terms on the $D_{SL}$ formula simply encode the evaluation in shifted variables of lower order impulses. 

This analysis readily extends to the more generic case considered in Ref. \cite{Bern:2023ity}, where a relaxation of the spin supplementary condition results in extra degrees of freedom, which are encoded in the boost vector $K^i$. In that case, the operator $\mathbb{O}$ can stand for $K$ as well, and we need to generalise the algebra as
\begin{align}
    \{S_{i},S_{j}\}=\epsilon_{ijk} S_{k}\,,
\quad 
\{S_{i},K_{j}\}=\epsilon_{ijk} K_{k}\,,
\quad
\{K_{i},K_{j}\}=-\epsilon_{ijk} S_{k}\,,
\end{align}
	while promoting the $D_{SL}$ operator to
 \begin{align}
 \label{defDL}
    D_L(f,g)\equiv -\epsilon_{ijk}\bigg(S_{k}\frac{\pd f}{\pd S_{i}}
  + K_{k}\frac{\pd f}{\pd K_{i}}\bigg)\frac{\pd g}{\underline{\pd L^j}}\, .
\end{align}
The crucial point is that due to the generalisation of the algebra, we now have 
\begin{align}
		\{\chi_1, \Delta \mathbb{O}^{(1),i}\} =
		\frac{\pd \Delta \mathbb{O}^{(1),i}}{\pd S^j}
		\{\chi_1, S^j\} = 
		\frac{\pd \Delta \mathbb{O}^{(1),i}}{\pd S^j}\Delta S^{(1),j}+\frac{\pd \Delta \mathbb{O}^{(1),i}}{\pd K^j}\Delta K^{(1),j}.
	\end{align}
Furthermore, we can again consider
\begin{align}
		D_{L}(\chi_1,g)&= \{S^j,\chi_1\} \frac{\pd g}{\underline{\pd L^j}}
		= \D L^{(1),j} \frac{\pd g}{\underline{\pd L^j}},
	\end{align}
where the two derivatives in eq. (\ref{defDL}) correspond to the two brackets in the algebra involving $S^i$. The rest of the derivation follows in analogy, and results in 
\begin{align*}
		\Delta \mathbb{O}_{\textrm{iter}}^{(2),i}=&
		\frac{1}{2}\frac{\pd \D \mathbb{O}^{(1),i}}{\pd S^j}\D S^{(1),j}+\frac{1}{2}\frac{\pd \D \mathbb{O}^{(1),i}}{\pd K^j}\D K^{(1),j}
		+\frac{1}{2} \frac{\pd \D \mathbb{O}^{(1),i}}{\pd L^j} \D L^{(1),j}
		+\frac{1}{2} \frac{\pd \D \mathbb{O}^{(1),i}}{\pd p^j} \D p^{(1),j}\,,
	\end{align*}
again showing the iterations simply come from the evaluation in shifted variables.
We have verified the validity of this formula to second order in the coupling and linear order in either $S^i$ or $K^i$, for the impulse, spin kick and boost kick, as in Ref. \cite{Bern:2023ity}.
	
	\section{Conclusions}
 \label{section:Conclusions}

In this paper, we studied the classical dynamics of spinning particles using scattering amplitudes and eikonal resummation.
A key assumption of our work was
that the loop expansion of scattering amplitudes can be rewritten as an exponential.
This eikonal exponentiation is by now well understood for scalar particles
(see~\cite{DiVecchia:2023frv} for a comprehensive review). 
However the spinning case is more complicated, since the relevant amplitudes are matrices in spin space.
Consequently, the exponential form of the amplitudes is also a matrix.
It would be particularly useful to deepen our understanding of eikonal exponentation for spinning amplitudes.

Having assumed eikonal exponentiation, our path to determining observables followed standard methods. 
We chose an initial state in the domain of validity of the classical theory. 
Time evolution was conveniently captured using the $S$ matrix. 
Physical observables were determined as expectations of appropriate operators n the final state.
This is very much in the spirit of KMOC~\cite{Kosower:2018adc,Maybee:2019jus} and other work on observables and the eikonal~\cite{DiVecchia:2021bdo,Cristofoli:2021jas,DiVecchia:2022nna,DiVecchia:2022piu,Gatica:2023iws,Georgoudis:2023eke}.
As emphasised in~\cite{Cristofoli:2021jas,Georgoudis:2023eke}, care is required in the use of eikonal methods to determine the final state in a scattering event because the delta-function support of the eikonal differs from the actual on-shell condition for final states.
In this paper, we introduced a systematic way of accounting for this mismatch involving a translation operator in momentum space.
This translation shifts momenta in the eikonal such that the correct, physical, on-shell condition is satisfied.

The translation led directly to an interesting structure of the eikonal function for spinning particles. 
A particularly important point is that, after translation and at the order we studied, the classical eikonal organises itself around the ``half-way'' point of the scattering event.
That is, the translated eikonal depends on $p + \Delta p/2$: the momentum plus half the total impulse,
rather than on a particle's momentum $p$ alone.
Similarly the classical eikonal function depends on the ``half-way'' spin rather than initial or final spins.

A beautiful aspect of the eikonal approach to classical dynamics is that the actual values of observables follow from stationary phase conditions.
One of these stationary phase equations picks a specific spatial point $x$, which is the impact parameter $b$ at lowest order.
In the case of spin, we found the determination of $x$ to be particularly subtle. 
It was very important to retain the matrix structure of the amplitude in this step.
In many respects, our approach is similar to the worldline approach of reference~\cite{Jakobsen:2021zvh}. 
However, in the worldline approach the position was fixed by demanding conservation of angular momentum.
We note that the position appearing in the eikonal in our approach is in general different to the position relevant for conservation of angular momentum. 
The two points happen to agree at the accuracy we required in this paper, but in general we expect these points to differ at higher orders.

Our approach correctly reproduced known expressions for the impulse and spin kick at one loop, and to quadratic order in spin. 
These observables are known to have a complicated structure, but in our approach this complexity arose from an algorithm which is very simple. 
Observables become complicated because as one continues to higher orders, one has to recall that lower order expressions for the observable involved positions, momenta and spins which must all be corrected to go to the next order in accuracy.
This iterative structure generates complex expressions from simple rules.
Clearly it will be interesting to study these ideas at higher orders (both in terms of loops and spins).
We expect the same basic ideas to apply, though considerable care will be required to account for effects which we were able to neglect.

In the context of the spin kick, we imposed the constraint of equation~\eqref{eq:constraint}.
Constraints have also arisen in other approaches to spin, notably reference~\cite{Bern:2020buy}.
Clarifying the origin of these constraints would be very useful for further connecting scattering amplitudes with classical physics.
Constraints are very well understood classically; evidently they must also emerge from quantum field theory but how this happens remains unclear.

The literature already contains a number of papers examining amplitudes, observables and classical physics using eikonal methods.
We showed how our approach is consistent with the canonical approach of Ref.~\cite{Bern:2020buy}  and the worldline formalism of Ref.~\cite{Jakobsen:2021zvh}.
Assuming that our assumption of eikonal exponentiation is valid, our approach opens the way towards an all-order understanding of spinning observables in the contect of the eikonal.

It is perhaps worth observing that the KMOC formalism~\cite{Kosower:2018adc,Maybee:2019jus} already provides an amplitudes-based method of computing classical observables for spinning particles which is expected to hold to all orders in perturbation theory. 
However eikonal exponentiation may provide a way to determine observables at large scattering angle; the KMOC formalism (as currently understood) is intrinsically perturbative in the scattering angle.

Much work remains to be performed before we can really say that we understand how spin, amplitudes, and classical physics work together in detail.
We completely neglected radiation, for example. 
It would be very interesting to understand how radiation and spin fit together in a more general eikonal picture.
As in the case of scalar particles, radiation requires incorporating the state of the radiation field, presumably as a coherent-state factor along the lines of references~\cite{Ciafaloni:2018uwe,Cristofoli:2021jas,DiVecchia:2022nna,DiVecchia:2022piu}.
A more general eikonal along these lines could be very useful for understanding spin effects in gravitational waveforms which have very recently been studied using KMOC~\cite{Kosower:2018adc,Cristofoli:2021vyo} in references~\cite{DeAngelis:2023lvf,Brandhuber:2023hhl,Aoude:2023dui}.
We look forward to working on this topic in the near future.

\subsection*{Acknowledgments}
We thank Giulia Isabella, Guanda Lin, Jan Steinhoff, Fei Teng, Justin Vines and Mao Zeng for useful discussions.
A.~L. is supported by funds from the European Union’s Horizon 2020 research and innovation program under the Marie Sklodowska-Curie grant agreement No.~847523 ‘INTERACTIONS’. The work of NM was supported by the Science and Technology Facilities Council (STFC) Consolidated Grants ST/T000686/1 “Amplitudes, Strings \& Duality” and ST/X00063X/1 “Amplitudes, Strings \& Duality”. No new data were generated or analysed during this study. DOC is supported in part by the STFC grant ``Particle Physics at the Higgs Centre''.
We are also grateful to the organisers of the ``Amplifying gravity at all scales'' program at Nordita, where parts of this project were undertaken.
 
 \appendix
	
	\section{Observables}
 \label{app:Observables}
We now present results for observables up to the 2PM approximation, and up to quadratic order in the spin of the particles. We may take as starting point an eikonal phase of the form
	\begin{align}
		\label{chii}
		\tilde{\chi}_n=& 
		\frac{\zeta_n}{4m_1m_2\sqrt{\gamma^2-1}} \,
		\, 
		\Big\{
		\alpha_n^{(0,1)}
		\log^{n-2}{|x_\perp|^2}+
		i\alpha_n^{(1,1)}
		\frac{\tilde{L}\cdot \tilde{a}_1}{|x_\perp|^2}+
		i\alpha_n^{(1,2)}
		\frac{\tilde{L}\cdot \tilde{a}_2}{|x_\perp|^2}\\
		\nonumber
		+&{\alpha'}_n^{(2,1)}
		\frac{(\tilde{L}\cdot \tilde{a}_1)(\tilde{L}\cdot \tilde{a}_2)}{|x_\perp|^4}+
		\alpha_n^{(2,2)}
		\frac{(\tilde{u}_2\cdot \tilde{a}_1)  (\tilde{u}_1\cdot \tilde{a}_2)}{|x_\perp|^2}+
		{\alpha'}_n^{(2,3)}
		\frac{(x_\perp\cdot \tilde{a}_1)(x_\perp\cdot \tilde{a}_2)}{|x_\perp|^4}
		\\
  \nonumber
		+&
		{\alpha'}_n^{(2,4)}
		\frac{(\tilde{L}\cdot \tilde{a}_1)^2}{|x_\perp|^4}+
		\alpha_n^{(2,5)}
		\frac{(\tilde{u}_2\cdot \tilde{a}_1)^2}{|x_\perp|^2}+
		{\alpha'}_n^{(2,6)}
		\frac{(x_\perp\cdot \tilde{a}_1)^2}{|x_\perp|^4}
		\Big\}
		\,,
	\end{align}
which can be understood as coming from a Fourier transform of a scattering amplitude (shown below), and whose kinematic prefactor comes from the integration over the two directions tansverse to the scattering plane. Here $|x_\perp|=\sqrt{-x_\perp^2}$, and the normalisation factors in position space $\zeta$ are given by 
	\begin{align}
		\zeta_1=-2G,\qquad\qquad \zeta_2=\frac{\pi G^2}{|x_\perp|},
	\end{align}
and we use the vector
\begin{equation}
		\tilde{L}^\mu \equiv \varepsilon^{\mu}(\tilde{u}_1,\tilde{u}_2,x_\perp).
	\end{equation}
The coefficients $\alpha'$ are given by
\begin{align*}
    {\alpha'}_1^{(2,1)}=&
    -{\alpha'}_1^{(2,3)}=
    \alpha_1^{(2,1)}-(\gamma^2-1)\alpha_1^{(2,3)},\\ {\alpha'}_2^{(2,1)}=&-2\alpha_2^{(2,1)}+(\gamma^2-1)\alpha_2^{(2,3)},\quad
 {\alpha'}_2^{(2,3)}=\alpha_2^{(2,1)}-2(\gamma^2-1)\alpha_2^{(2,3)},\\
    {\alpha'}_1^{(2,4)}=&-{\alpha'}_1^{(2,6)}=\alpha_1^{(2,4)}-(\gamma^2-1)\alpha_1^{(2,6)},\\ {\alpha'}_2^{(2,4)}=&-2\alpha_2^{(2,4)}+(\gamma^2-1)\alpha_2^{(2,6)},\quad
{\alpha'}_2^{(2,6)}=\alpha_2^{(2,4)}-2(\gamma^2-1)\alpha_2^{(2,6)}.
\end{align*}
As discussed in the text, we perform a shift in impact parameter in the Eikonal phase 
	\begin{equation}
		x_\perp^\mu \rightarrow x_\perp^\mu + \xi_2 \tilde{u}_2^\mu - \xi_1 \tilde{u}_1^\mu,
	\end{equation}
	where 
	\begin{equation}
		\xi_1=\frac{x_\perp\cdot (\gamma \tilde{u}_2 - \tilde{u}_1)}{\gamma^2-1},
		\qquad
		\xi_2=\frac{x_\perp\cdot (\tilde{u}_2 - \gamma \tilde{u}_1)}{\gamma^2-1}.
	\end{equation}
	
	For the 2PM results, we split it in general as
	\begin{align}
		\D \mathbb{O}^{(i,j)}
		=
		\frac{G}{4m_1m_2\sqrt{\gamma^2-1}|b|^{2(\zeta_{\mathbb{O}}+i+j)}}
		\bigg\{ &
		\sum_k 
		\mathcal{O}_{\Delta \mathbb{O},\chi}^{(i,j,k)}
		\Big(
		c_{\Delta \mathbb{O},\chi_1}^{(i,j,k)}+
		\frac{G\pi}{|b|}c_{\Delta \mathbb{O},\chi_2}^{(i,j,k)}
		\Big)
		\\ \nonumber
		&-
		\frac{G}{2m_1^2m_2^2(\gamma^2-1)^{3/2}}
		\sum_l
		\mathcal{O}_{\Delta \mathbb{O},\textrm{iter}}^{(i,j,l)}
		c_{\Delta \mathbb{O},\textrm{iter}}^{(i,j,l)}
		\bigg\},
	\end{align}
	where $\mathbb{O}$ means either $p$ or $a_1$ for the impulse or the spin kick, respectively, and the observable-dependent factor $\zeta_{\mathbb{O}}$ is given by
	\begin{align}
		\zeta_{p}=1,
		\qquad\qquad 
		\zeta_{a_1}=0.
	\end{align}
	The indices $i$ and $j$ label the powers of $a_1$ and $a_2$, respectively. The structures $\mathcal{O}_{\Delta p,\chi}$ correspond to contributions linear in the Eikonal phase at either the first or second post Minkowskian approximation, and so the coefficients $c_{\Delta p,\chi_i}$ depend on one $\alpha_i$, while the $\mathcal{O}_{\Delta p,\textrm{iter}}$ correspond to iterations, and their coefficients $c_{\Delta p,\textrm{iter}}$ depend on products of two $\alpha_1$'s.

 \newpage
	The structures for $\Delta p$ and $\Delta a_1$ which contribute linearly in the Eikonal phase are
	
	%\begin{table}[h!]
	%\centering
	\begin{tabular}{ll}
		$\mathcal{O}_{\Delta p,\chi}^{(0,0,1)}=b^\mu$   &  
		$\mathcal{O}_{\Delta a_1,\chi}^{(1,0,1)}=(a_1\cdot u_2)b^\mu$\\[4pt]
		$\mathcal{O}_{\Delta p,\chi}^{(1,0,1)}=(a_1\cdot L)b^\mu$   &  
		$\mathcal{O}_{\Delta a_1,\chi}^{(1,0,2)}=(a_1\cdot b)u_1^\mu$\\[4pt]
		$\mathcal{O}_{\Delta p,\chi}^{(1,0,2)}=(a_1\cdot b)L^\mu$     & 
		$\mathcal{O}_{\Delta a_1,\chi}^{(1,0,3)}=(a_1\cdot b)u_2^\mu$\\[4pt]
		$\mathcal{O}_{\Delta p,\chi}^{(1,1,1)}=(\gamma^2-1)(a_1\cdot b)(a_2\cdot b)b^\mu$   &  
		$\mathcal{O}_{\Delta a_1,\chi}^{(1,1,1)}=(a_2\cdot L)\big((a_1\cdot b)u_2^\mu-(a_1\cdot u_2)b^\mu\big)$\\[4pt]
		$\mathcal{O}_{\Delta p,\chi}^{(1,1,2)}=\big((a_1\cdot b)(a_2\cdot L)+(a_1\cdot L)(a_2\cdot b)\big)L^\mu$   &  
		$\mathcal{O}_{\Delta a_1,\chi}^{(1,1,2)}=(a_2\cdot b)\big((a_1\cdot u_2)L^\mu-(a_1\cdot L)u_2^\mu\big)$\\[4pt]
		$\mathcal{O}_{\Delta p,\chi}^{(1,1,3)}=(a_1\cdot L)(a_2\cdot L)b^\mu$     & 
		$\mathcal{O}_{\Delta a_1,\chi}^{(1,1,3)}=(a_2\cdot u_1)\big((a_1\cdot b)L^\mu-(a_1\cdot L)b^\mu\big)$\\[4pt]
		$\mathcal{O}_{\Delta p,\chi}^{(1,1,4)}=b^2(a_1\cdot u_2)(a_2\cdot u_1)b^\mu$   &  
		$\mathcal{O}_{\Delta a_1,\chi}^{(1,1,4)}=(a_2\cdot b)(a_1\cdot L)u_1^\mu$\\[4pt]
		$\mathcal{O}_{\Delta p,\chi}^{(2,0,1)}=(\gamma^2-1)(a_1\cdot b)^2b^\mu$   &  
		$\mathcal{O}_{\Delta a_1,\chi}^{(1,1,5)}=(a_1\cdot b)(a_2\cdot L)u_1^\mu$\\[4pt]
		$\mathcal{O}_{\Delta p,\chi}^{(2,0,2)}=(a_1\cdot b)(a_1\cdot L)L^\mu$      & 
		$\mathcal{O}_{\Delta a_1,\chi}^{(2,0,1)}=(a_1\cdot u_2)(a_1\cdot L)b^\mu$\\[4pt]
		$\mathcal{O}_{\Delta p,\chi}^{(2,0,3)}=(a_1\cdot L)^2b^\mu$   &  
		$\mathcal{O}_{\Delta a_1,\chi}^{(2,0,2)}=(a_1\cdot u_2)(a_1\cdot b)L^\mu$\\[4pt]
		$\mathcal{O}_{\Delta p,\chi}^{(2,0,4)}=b^2(a_1\cdot u_2)^2b^\mu$   &  
		$\mathcal{O}_{\Delta a_1,\chi}^{(2,0,3)}=(a_1\cdot b)(a_1\cdot L)u_1^\mu$\\[4pt]
		& 
		$\mathcal{O}_{\Delta a_1,\chi}^{(2,0,4)}=(a_1\cdot b)(a_1\cdot L)u_2^\mu$\\[4pt]
	\end{tabular}
	%\end{table} 
	
	The 1PM coefficients for $\Delta p$ are
	\begin{align*}    
		c_{\Delta p,\chi_1}^{(0,0,1)}&=-2\alpha_1^{(0,1)}\\
		c_{\Delta p,\chi_1}^{(1,0,1)}&=c_{\Delta p,\chi_1}^{(1,0,2)}= 2i\alpha_1^{(1,1)},\\ 
		-c_{\Delta p,\chi_1}^{(1,1,1)}&=
		c_{\Delta p,\chi_1}^{(1,1,2)}=
		c_{\Delta p,\chi_1}^{(1,1,3)}=
		4(\alpha_1^{(2,1)}-\alpha_1^{(2,3)}),
		\qquad
		c_{\Delta p,\chi_1}^{(1,1,4)}=0,
		\\
		-2c_{\Delta p,\chi_1}^{(2,0,1)}&=
		c_{\Delta p,\chi_1}^{(2,0,2)}=
		2c_{\Delta p,\chi_1}^{(2,0,3)}=8(\alpha_1^{(2,4)}-\alpha_1^{(2,6)}),
		\qquad
		c_{\Delta p,\chi_1}^{(2,0,4)}=0,
	\end{align*}
	while the 1PM coefficients for $\Delta a_1$ are
	\begin{align*}  
		c_{\Delta a_1,\chi_1}^{(1,0,1)}&=
		-c_{\Delta a_1,\chi_1}^{(1,0,3)}=
		2i\alpha_1^{(1,1)},\\
		c_{\Delta a_1,\chi_1}^{(1,0,2)}&=\gamma c_{\Delta a_1,\chi_1}^{(1,0,1)}+2\alpha_1^{(0,1)}\\
		c_{\Delta a_1,\chi_1}^{(1,1,1)}&=
		-c_{\Delta a_1,\chi_1}^{(1,1,2)}=-2\alpha_1^{(2,1)}+2\alpha_1^{(2,3)}\\
		c_{\Delta a_1,\chi_1}^{(1,1,3)}&=0\\
		c_{\Delta a_1,\chi_1}^{(1,1,4)}&=
		c_{\Delta a_1,\chi_1}^{(1,1,5)}=
		-2i\alpha_1^{(1,2)}-\gamma c_{\Delta a_1,\chi_1}^{(1,1,1)}\\
		c_{\Delta a_1,\chi_1}^{(2,0,1)}&=
		c_{\Delta a_1,\chi_1}^{(2,0,2)}=
		-c_{\Delta a_1,\chi_1}^{(2,0,4)}/2=
		4\alpha_1^{(2,4)}-4\alpha_1^{(2,6)}\\
		c_{\Delta a_1,\chi_1}^{(2,0,3)}&=
		2\gamma c_{\Delta a_1,\chi_1}^{(2,0,1)}-4i\alpha_1^{(1,1)}.
	\end{align*}
	
	\newpage
	The 2PM coefficients for both $\Delta p$ and $\Delta a_1$ are
	
	%\begin{table}[h!]
	%\centering
	\begin{tabular}{ll}
		$c_{\Delta p,\chi_2}^{(0,0,1)}=-\alpha_2^{(0,1)}$   &  
		$c_{\Delta a_1,\chi_2}^{(1,0,1)}=i\alpha_2^{(1,1)}$\\[4pt]
		$c_{\Delta p,\chi_2}^{(1,0,1)}=2i\alpha_2^{(1,1)}$   &  
		$c_{\Delta a_1,\chi_2}^{(1,0,2)}=\alpha_2^{(0,1)}+\gamma  i\alpha_2^{(1,1)}$\\[4pt]
		$c_{\Delta p,\chi_2}^{(1,0,2)}=i\alpha_2^{(1,1)}$   &  
		$c_{\Delta a_1,\chi_2}^{(1,0,3)}=-i\alpha_2^{(1,1)}$\\[4pt]
		$c_{\Delta p,\chi_2}^{(1,1,1)}=-3\alpha_2^{(2,1)}+6\alpha_2^{(2,3)}$       &  
		$c_{\Delta a_1,\chi_2}^{(1,1,1)}=-2\alpha_2^{(2,1)}+\alpha_2^{(2,3)}$\\[4pt]
		$c_{\Delta p,\chi_2}^{(1,1,2)}=3\alpha_2^{(2,1)}-3\alpha_2^{(2,3)}$     &  
		$c_{\Delta a_1,\chi_2}^{(1,1,2)}=\alpha_2^{(2,1)}-2\alpha_2^{(2,3)}$\\[4pt]
		$c_{\Delta p,\chi_2}^{(1,1,3)}=6\alpha_2^{(2,1)}-3\alpha_2^{(2,3)}$   &  
		$c_{\Delta a_1,\chi_2}^{(1,1,3)}=-\alpha_2^{(2,2)}$\\[4pt]
		$c_{\Delta p,\chi_2}^{(1,1,4)}=3\alpha_2^{(2,2)}$   &  
		$c_{\Delta a_1,\chi_2}^{(1,1,4)}=-i\alpha_2^{(1,2)}+\gamma c_{\Delta a_1,\chi_2}^{(1,1,2)}$\\[4pt]
		$c_{\Delta p,\chi_2}^{(2,0,1)}=-3\alpha_2^{(2,4)}+6\alpha_2^{(2,6)}$   &  
		$c_{\Delta a_1,\chi_2}^{(1,1,5)}=-2i\alpha_2^{(1,2)}-\gamma c_{\Delta a_1,\chi_2}^{(1,1,1)}$\\[4pt]
		$c_{\Delta p,\chi_2}^{(2,0,2)}=6\alpha_2^{(2,4)}-6\alpha_2^{(2,6)}$   &  
		$c_{\Delta a_1,\chi_2}^{(2,0,1)}=4\alpha_2^{(2,4)}+2\alpha_2^{(2,5)}-2\alpha_2^{(2,6)}$\\[4pt]
		$c_{\Delta p,\chi_2}^{(2,0,3)}=6\alpha_2^{(2,4)}-3\alpha_2^{(2,6)}$   &  
		$c_{\Delta a_1,\chi_2}^{(2,0,2)}=2\alpha_2^{(2,4)}-2\alpha_2^{(2,5)}-4\alpha_2^{(2,6)}$\\[4pt]
		$c_{\Delta p,\chi_2}^{(2,0,4)}=3\alpha_2^{(2,5)}$   &  
		$c_{\Delta a_1,\chi_2}^{(2,0,3)}=-3i\alpha_2^{(1,1)}-\gamma c_{\Delta a_1,\chi_2}^{(2,0,4)}$\\[4pt]
		&  
		$c_{\Delta a_1,\chi_2}^{(2,0,4)}=-6\alpha_2^{(2,4)}+6\alpha_2^{(2,6)}$\\[4pt]
	\end{tabular}
	%\end{table} 
	
	The iteration structures are
	
	%\begin{table}[h!]
	%\centering
	\begin{tabular}{ll}
		$\mathcal{O}_{\Delta p,\textrm{iter}}^{(0,0,1)}=K^\mu$   &  
		$\mathcal{O}_{\Delta a_1,\textrm{iter}}^{(1,0,1)}=b^{-2}(a_1\cdot b)b^\mu$\\[4pt]
		$\mathcal{O}_{\Delta p,\textrm{iter}}^{(1,0,1)}=(a_1\cdot u_2)L^\mu$   &  
		$\mathcal{O}_{\Delta a_1,\textrm{iter}}^{(1,0,2)}=(a_1\cdot u_2)u_1^\mu$\\[4pt]
		$\mathcal{O}_{\Delta p,\textrm{iter}}^{(1,0,2)}=(a_1\cdot L)K^\mu$   &  
		$\mathcal{O}_{\Delta a_1,\textrm{iter}}^{(1,0,3)}=(a_1\cdot u_2)u_2^\mu$\\[4pt]
		$\mathcal{O}_{\Delta p,\textrm{iter}}^{(1,1,1)}=(a_2\cdot u_1)(a_1\cdot t)^\mu$       &  
		$\mathcal{O}_{\Delta a_1,\textrm{iter}}^{(1,1,1)}=(a_2\cdot u_1)\big((a_1\cdot u_2)L^\mu-(a_1\cdot L)u_2^\mu\big)$\\[4pt]
		$\mathcal{O}_{\Delta p,\textrm{iter}}^{(1,1,2)}=(a_1\cdot u_2)(a_2\cdot t)^\mu$     &  
		$\mathcal{O}_{\Delta a_1,\textrm{iter}}^{(1,1,2)}=b^{-2}(a_2\cdot b)\big((a_1\cdot b)L^\mu-(a_1\cdot L)b^\mu\big)$\\[4pt]
		$\mathcal{O}_{\Delta p,\textrm{iter}}^{(1,1,3)}=(\gamma^2-1)(a_1\cdot b)(a_2\cdot b)K^\mu$   &  
		$\mathcal{O}_{\Delta a_1,\textrm{iter}}^{(1,1,3)}=b^{-2}(a_1\cdot b)\big((a_2\cdot b)L^\mu+(a_2\cdot L)b^\mu\big)$\\[4pt]
		$\mathcal{O}_{\Delta p,\textrm{iter}}^{(1,1,4)}=(a_1\cdot L)(a_2\cdot L)K^\mu$   &  
		$\mathcal{O}_{\Delta a_1,\textrm{iter}}^{(1,1,4)}=(a_1\cdot u_2)(a_2\cdot L)u_2^\mu$\\[4pt]
		$\mathcal{O}_{\Delta p,\textrm{iter}}^{(2,0,1)}=(a_1\cdot u_2)(a_1\cdot t)^\mu$   &  
		$\mathcal{O}_{\Delta a_1,\textrm{iter}}^{(1,1,5)}=(a_2\cdot u_1)(a_1\cdot L)u_1^\mu$\\[4pt]
		$\mathcal{O}_{\Delta p,\textrm{iter}}^{(2,0,2)}=(\gamma^2-1)(a_1\cdot b)^2 K^\mu$   &  
		$\mathcal{O}_{\Delta a_1,\textrm{iter}}^{(1,1,6)}=(a_1\cdot u_2)(a_2\cdot L)u_1^\mu$\\[4pt]
		$\mathcal{O}_{\Delta p,\textrm{iter}}^{(2,0,3)}=(a_1\cdot L)^2 K^\mu$   &  
		$\mathcal{O}_{\Delta a_1,\textrm{iter}}^{(2,0,1)}=(a_1\cdot u_2)^2L^\mu$\\[4pt]
		&  
		$\mathcal{O}_{\Delta a_1,\textrm{iter}}^{(2,0,2)}=b^{-2}(a_1\cdot b)(a_1\cdot L)b^\mu$ \\[4pt]
		&  
		$\mathcal{O}_{\Delta a_1,\textrm{iter}}^{(2,0,3)}=b^{-2}(a_1\cdot b)^2 L^\mu$\\[4pt]
		&  
		$\mathcal{O}_{\Delta a_1,\textrm{iter}}^{(2,0,4)}=(a_1\cdot u_2)(a_1\cdot L)u_2^\mu$\\[4pt]
		&  
		$\mathcal{O}_{\Delta a_1,\textrm{iter}}^{(2,0,5)}=(a_1\cdot u_2)(a_1\cdot L)u_1^\mu$
	\end{tabular}
	%\end{table} 
	
	where we have used the definitions
	\begin{align}
 K^\mu&=(m_1+\gamma m_2)u_2^\mu -(m_2+\gamma m_1)u_1^\mu\\
		(a_i\cdot t)^\mu &\equiv {a_i}_\lambda t^{\lambda\mu},\qquad\qquad
		t^{\lambda\mu} \equiv (\gamma^2-1)b^\lambda b^\mu-3L^\lambda L^\mu.
	\end{align}
	
	The iteration coefficients for the impulse are
	\begin{align*}
		c_{\Delta p,\textrm{iter}}^{(0,0,1)} & = -g_ {0, 0}\\
		c_{\Delta p,\textrm{iter}}^{(1,0,1)} & = 
		2 (m_ 1 + m_ 2 \gamma) g_ {0, 1} + 
		m_ 2 (g_ {0, 0} + (-1 + \gamma^2) g_ {1, 1}) \\
		c_{\Delta p,\textrm{iter}}^{(1,0,2)} & = 2 g_ {0, 1} \\
		c_{\Delta p,\textrm{iter}}^{(1,1,1)} & = -(m_ 2 + m_ 1 \gamma) (g_ {1, 2} + g_ {0, 3}) - 
		m_ 1 (g_ {0, 1} + (-1 + \gamma^2) g_ {2, 3}) \\
		c_{\Delta p,\textrm{iter}}^{(1,1,2)} & = (m_ 1 + m_ 2 \gamma) ( g_ {1, 2} + g_ {0, 3}) + 
		m_ 2 (g_ {0, 2} + (-1 + \gamma^2) g_ {1, 3}) \\
		c_{\Delta p,\textrm{iter}}^{(1,1,3)} & = -2 (g_ {1, 2} - 2 g_ {0, 3}), \qquad
		c_{\Delta p,\textrm{iter}}^{(1,1,4)} = -2 (g_ {1, 2} + 2 g_ {0, 3}) \\
		c_{\Delta p,\textrm{iter}}^{(2,0,1)} & = (m_ 1 + m_ 2 \gamma) ( g_ {1, 1} + 2 g_ {0, 4} ) + 
		m_ 2 (g_ {0, 1} + 2 (-1 + \gamma^2) g_ {1, 4}) \\
		c_{\Delta p,\textrm{iter}}^{(2,0,2)} & = -(g_ {1, 1} - 4 g_ {0, 4}), \qquad
		c_{\Delta p,\textrm{iter}}^{(2,0,3)} = -(g_ {1, 1} + 4 g_ {0, 4})
\end{align*}
  
  The iteration coefficients for the spin kick are
	\begin{align*}
  c_{\Delta a_1,\textrm{iter}}^{(1,0,1)} & = 
		m_ 2 (-1 + \gamma^2) (-g_ {0, 0} + (-1 + \gamma^2) g_ {1, 1}) \\
		c_{\Delta a_1,\textrm{iter}}^{(1,0,2)} & = 
		m_ 2 (-1 + \gamma^2) ( 
		2 g_ {0, 1} + \gamma g_ {1, 1}) + (m_ 1 + m_ 2 \gamma) g_ {0, 0} \\
		c_{\Delta a_1,\textrm{iter}}^{(1,0,3)} & = m_ 2 (-1 + \gamma^2) (-g_ {1, 1}) \\
		c_{\Delta a_1,\textrm{iter}}^{(1,1,1)} & =-
		c_{\Delta p,\textrm{iter}}^{(1,1,1)}\\
		c_{\Delta a_1,\textrm{iter}}^{(1,1,2)} & = 
		m_ 2 (-1 + \gamma^2) (-g_ {0, 2} + (-1 + \gamma^2) g_ {1, 
			3}) + (m_ 1 + m_ 2 \gamma) (-1 + \gamma^2) (g_ {1, 2} - 
		g_ {0, 3}) \\
		c_{\Delta a_1,\textrm{iter}}^{(1,1,3)} & = 
		m_ 2 (-1 + \gamma^2) (2 g_ {0, 2} - 2 (-1 + \gamma^2) g_ {1, 3}), \qquad
		c_{\Delta a_1,\textrm{iter}}^{(1,1,4)}  = m_ 2 (-1 + \gamma^2) (2 g_ {1, 3}) \\
		c_{\Delta a_1,\textrm{iter}}^{(1,1,5)} & = 
		m_ 1 (g_ {0, 0} + \gamma   g_ {0, 
			1} + (-1 + \gamma^2) (g_ {2, 2} + \gamma g_ {2, 3})) + (m_ 2 + 
		m_ 1 \gamma)  (2 g_ {0, 2} + \gamma g_ {0, 3} + \gamma g_ {1, 
			2}) \\
		c_{\Delta a_1,\textrm{iter}}^{(1,1,6)} & = 
		m_ 2 (-1 + \gamma^2) (-2 g_ {1, 2} - 2 g_ {0, 3} - 
		2\gamma  g_ {1, 3}) + (m_ 1 + m_ 2 \gamma) (-2 g_ {0, 2}) \\
		c_{\Delta a_1,\textrm{iter}}^{(2,0,1)} & =
		-c_{\Delta p,\textrm{iter}}^{(2,0,1)}\\
		c_{\Delta a_1,\textrm{iter}}^{(2,0,2)} & = 
		m_ 2 (-1 + \gamma^2) (-6  (-1 + \gamma^2) g_ {1, 4} + 
		3 g_ {0, 1}) + (m_ 1 + m_ 2 \gamma) (-1 + \gamma^2) ( 
		2  g_ {0, 4} - g_ {1, 1}) \\
		c_{\Delta a_1,\textrm{iter}}^{(2,0,3)} & = 
		m_ 2 (-1 + \gamma^2) (4 g_ {0, 1} - 8 (-1 + \gamma^2) g_ {1, 4})-c_{\Delta a_1,\textrm{iter}}^{(2,0,2)} \\
		c_{\Delta a_1,\textrm{iter}}^{(2,0,4)} & = 4 m_ 2 g_ {0, 1} + 4 (m_ 1 + m_ 2 \gamma) g_ {0, 4} - (-1+\gamma^2)^{-1} c_{\Delta a_1,\textrm{iter}}^{(2,0,2)} \\
		c_{\Delta a_1,\textrm{iter}}^{(2,0,5)} & = 
		m_ 2 (-1 + \gamma^2) ( (-6\gamma g_ {1, 4} - 6 g_ {0, 4} - 
		3 g_ {1, 1})) + 
		m_ 2 (-\gamma g_ {0, 1} - g_ {0, 0})\\
		&\qquad\qquad\qquad\qquad - (m_2+m_1 \gamma) (2 g_ {0, 4}) - (m_ 1 + 
		m_ 2 \gamma)  (4  g_ {0, 1} + \gamma g_ {1, 1}),
	\end{align*}
where
\begin{align}
    g_{i,j}&\equiv \beta_i \beta_j,\\
    \nonumber 
    \beta_0&=\alpha_1^{(0,1)},\qquad 
    \beta_1=i\alpha_1^{(1,1)},\qquad
    \beta_2=i\alpha_1^{(1,2)},\\
    \beta_3&=-\alpha_1^{(2,1)}+\alpha_1^{(2,3)},\qquad
    \beta_4=-\alpha_1^{(2,4)}+\alpha_1^{(2,6)}.
\end{align}
The $\alpha_n^{(i,j)}$ coefficients are those appearing in a scattering amplitude of the form
	\begin{align}
		\label{MnPM}
		\mathcal{M}^{\textrm{nPM}} = 
		\zeta_{q,n}
		\Big\{
		&\alpha_n^{(0,1)}+
		\alpha_n^{(1,1)}\cE_1+
		\alpha_n^{(1,2)}\cE_2\\
		\nonumber
		+&\alpha_n^{(2,1)}\cE_1\cE_2+
		\alpha_n^{(2,2)}q^2 (u_2\cdot a_1)  (u_1\cdot a_2)+
		(-1+\gamma^2)\alpha_n^{(2,3)}(q\cdot a_1)(q\cdot a_2)
		\\
		\nonumber
		+&
		\alpha_n^{(2,4)}\cE_1^2\ \ \,
		+\alpha_n^{(2,5)}q^2(u_2\cdot a_1)^2
		\qquad\quad\,
		+(-1+\gamma^2)\alpha_n^{(2,6)}(q\cdot a_1)^2
		\Big\}
		\,,
	\end{align}
	where the order-dependent factors $\zeta_n$ are given by
	\begin{align}
		\zeta_{q,1}=-\frac{4\pi G}{q^2}, \qquad \zeta_{q,2}=\frac{2\pi^2 G^2}{\sqrt{-q^2}},
	\end{align}
	and $\cE_i\equiv \epsilon^{\mu\nu\rho\sigma}{u_1}_\mu {u_2}_\nu q_\rho a_1{}_\sigma$. These amplitudes are given in a manifestly shift-symmetric basis. For uniformity, we have included terms containing $q^2$ in the numerator also for the 1PM, but they produce local terms upon Fourier transform, and can be discarded.
	From such an amplitude, we can obtain an Eikonal phase like the one above through Fourier transform.
	In the case of gravity, the tree-level coefficients $\alpha_1^{(i,j)}$ take the explicit form
	\begin{align}
		\nonumber 
		\alpha_1^{(0,1)}&=
		4m_1^2m_2^2(2\gamma^2-1)
		\,,\qquad
		\alpha_1^{(1,1)}=
		-8im_1^2m_2^2\gamma
		\,,\qquad 
		\alpha_1^{(1,2)}=
		-8im_1^2m_2^2\gamma,
		\\
		\alpha_1^{(2,1)}&=
		\frac{4m_1^2m_2^2(1-2\gamma^2)}{\gamma^2-1}
		\,,\qquad
		\alpha_1^{(2,4)}=
		\frac{2m_1^2m_2^2(2\gamma^2-1)(C_E-1)}{\gamma^2-1}\,,
		\\
		\nonumber 
		\alpha_1^{(2,2)}&=
		0
		\,,
		\,\,\qquad
		\alpha_1^{(2,3)}=
		0,
		\qquad 
		\alpha_1^{(2,5)}=
		0
		\,,\qquad 
		\alpha_1^{(2,6)}=
		0
		\,,
		%\label{eq:treeAlphas}
	\end{align}
	while for the one-loop amplitude, the coefficients $\alpha_2^{(i,j)}$ are given by
	\begin{align}
		\nonumber
		\alpha_2^{(0,1)}&=
		3m_1^2m_2^2(m_1+m_2)(5\gamma^2-1)
		\,,\\
		\nonumber 
		\alpha_2^{(1,1)}&=
		-\frac{im_1^2m_2^2(4m_1+3m_2)(5\gamma^2-3)\gamma}{(\gamma^2-1)}
		\,,\qquad 
		\alpha_2^{(1,2)}=
		-\frac{im_1^2m_2^2(3m_1+4m_2)(5\gamma^2-3)\gamma}{(\gamma^2-1)}
		\\
		\nonumber 
		\alpha_2^{(2,1)}&=
		-\frac{m_1^2m_2^2(m_1+m_2)(20\gamma^4-21\gamma^2+3)}{(\gamma^2-1)^2}
		\,,\qquad 
		\alpha_2^{(2,2)}=
		\frac{m_1^2m_2^2(m_1+m_2)(5\gamma^2-3)\gamma}{(\gamma^2-1)^2}
		\,,\\
		\nonumber
		\alpha_2^{(2,3)}&=0
		\,.
	\end{align}
	For the $S_1^2$ terms of the amplitude, we find it useful to split the coefficients as
	\begin{align}
		\alpha_2^{(2,i)}=
		\frac{m_1^2m_2^2}{16(\gamma^2-1)^2}
		\left(
		\alpha_{2,\textrm{BH}}^{(2,i)}
		+C_E\alpha_{2,C_E}^{(2,i)}\right),
		\qquad i=4,5,6.
	\end{align}
	where the Wilson coefficient $C_E$ vanishes for the black hole, and so the $\alpha_{2,\textrm{BH}}$ capture all the information in that limit. The pieces of the coefficient are given by
	\begin{align}
		\nonumber 
		\alpha_{2,\textrm{BH}}^{(2,4)}&=
		-2 \left(
		m_1\left(95 \gamma^4-102 \gamma^2+15\right)
		+4m_2\left(15 \gamma ^4-15 \gamma ^2+2\right)
		\right),
		\\
		\nonumber 
		\alpha_{2,C_E}^{(2,4)} &=\left(
		m_1\left(95 \gamma ^4-102\gamma ^2+23\right) +4 m_2\left(15\gamma ^4-13 \gamma ^2+2\right)
		\right)
		\,,\\
		\nonumber
		\alpha_{2,\textrm{BH}}^{(2,5)}&=
		-2 \left(
		m_1\left(35 \gamma^4-30 \gamma^2+3\right)
		-4m_2\left(3 \gamma ^2-1\right)
		\right),
		\\
		\nonumber 
		\alpha_{2,C_E}^{(2,4)}&=\left(
		m_1\left(35 \gamma ^4-30\gamma ^2+11\right) +4 m_2\left(3 \gamma ^2+1\right)
		\right)
		\,,\\
		\nonumber 
		\alpha_{2,C_E}^{(2,6)}&=
		-4
		\left(
		3m_1(5\gamma^2-1)+m_2(15\gamma^2-1)
		\right)
		\,\qquad \alpha_{2,\textrm{BH}}^{(2,6)}=0.
		\label{eq:1loopAlphas}
	\end{align}

	\section{Useful Expressions}
 \label{app:Useful}
	
 The Fourier transform is given by
 \[
	I^{\mu_1\mu_2\cdots\mu_{\beta}}\left(D,\alpha\right) & =\int \hat{\mathrm{d}}^D \hat{\delta}\left(2p_1\cdot q\right) \hat{\delta}\left(2p_2 \cdot q\right) e^{i q \cdot x} \frac{q^{\mu_1}q^{\mu_2}\cdots q^{\mu_\beta}}{q^{2\alpha}} \\
	& =(-i\frac{\pd}{\pd x_{\mu_1}})(-i\frac{\pd}{\pd x_{\mu_2}})\cdots (-i\frac{\pd}{\pd x_{\mu_\beta}})\frac{1}{4m_1m_2\sqrt{\gamma^2-1}} \frac{\Gamma(D / 2-1-\alpha)}{2^{2 \alpha} \pi^{D / 2-1} \Gamma(\alpha)} \frac{1}{|x|^{d-2-2 \alpha}}
	\]
	\[
	I\left(4-2\epsilon,1\right) = -\frac{1}{16\pi m_1m_2\sqrt{\gamma^2-1}}\log x^2 + \cl{O}(\frac{1}{\epsilon}) + const 
	\]
	\[
	I^\mu\left(4,1\right) = i\frac{1}{8\pi m_1m_2\sqrt{\gamma^2-1}}\frac{x^\mu}{x^2} 
	\]

	The values for $N_q, N_0$ and $N_{0q}$ are derived from the conditions $e_0^2 = -e_q^2 = 1$ and $e_q\cdot e_0 = 0$, which at order $G^0$ are given by
	\[
	N_0 &= \frac{1}{\sqrt{(p_1 + p_2)^2}} = \frac{1}{\sqrt{m_1^2 + m_2^2 + 2\gamma m_1m_2}},\\
	N_q &= \frac12\sqrt{\frac{(p_1 + p_2)^2}{((p_1\cdot p_2)^2 - p_1^2p_2^2)}} = \frac12\sqrt{\frac{m_1^2 + m_2^2 + 2\gamma m_1m_2}{m_1^2m_2^2(\gamma^2 - 1)}},\\
	N_{0q} &= -\frac12 \frac{m_1^2-m_2^2}{\sqrt{(p_1 + p_2)^2((p_1\cdot p_2)^2 - p_1^2p_2^2)}} = -\frac12 \frac{m_1^2-m_2^2}{\sqrt{(m_1^2 + m_2^2 + 2\gamma m_1m_2)(m_1^2m_2^2(\gamma^2 - 1))}}
	\]
	It is often useful to resolve certain vectors onto a basis $\{u_1^\mu, u_2^\mu, b^\mu, \varepsilon^\mu(b,u_1,u_2)\}$. A few useful expressions are
	\[
	\varepsilon^\mu(b,s,u_1) &= \frac{s\cdot u_2}{\gamma
		^2-1}\varepsilon^{\mu }\left(b,u_1,u_2\right) - \frac{\gamma \varepsilon \left(b,s,u_1,u_2\right)}{\gamma
		^2-1}u_1^{\mu } +\frac{\varepsilon \left(b,s,u_1,u_2\right)}{\gamma ^2-1}u_2^{\mu }\\
	\varepsilon^\mu(b,s,u_2) &= \frac{\gamma  s\cdot u_2}{\gamma ^2-1}\varepsilon^{\mu
	}\left(b,u_1,u_2\right)-\frac{ \varepsilon
		\left(b,s,u_1,u_2\right)}{\gamma ^2-1}u_1^{\mu }+\frac{\gamma  \varepsilon
		\left(b,s,u_1,u_2\right)}{\gamma ^2-1} u_2^{\mu }\\
	\varepsilon^\mu(s,u_1,u_2) &= \frac{b\cdot s }{b^2}\varepsilon^{\mu
	}\left(b,u_1,u_2\right)+\frac{\varepsilon
		\left(b,s,u_1,u_2\right)}{b^2}b^{\mu } 
	\]
	\bibliographystyle{JHEP}
	\bibliography{mainbib} 
\end{document}